\documentclass[aps,prd,twocolumn,superscriptaddress,showpacs,floatfix]{revtex4}
\usepackage{bm}
\usepackage{graphicx}

\newcommand{\xvec}{\bm{x}}

\newcommand{\pvec}{\bm{p}}

\begin{document}
\title{Group-theoretical construction of extended baryon operators in lattice QCD}
\author{S.~Basak}
\affiliation{Department of Physics, 
             University of Maryland,
             College Park, MD 20742, USA}
\author{R.G.~Edwards}
\affiliation{Thomas Jefferson National Accelerator Facility,
             Newport News, VA 23606, USA}
\author{G.T.~Fleming}
\affiliation{Sloane Physics Laboratory, 
             Yale University, 
             New Haven, CT 06520, USA}
\author{U.M.~Heller}
\affiliation{American Physical Society,
             Ridge, NY 11961-9000, USA}
\author{C.~Morningstar}
\affiliation{Department of Physics, 
             Carnegie Mellon University, 
             Pittsburgh, PA 15213, USA}
\author{D.~Richards}
\affiliation{Thomas Jefferson National Accelerator Facility,
             Newport News, VA 23606, USA}
\author{I.~Sato}
\author{S.~Wallace}
\affiliation{Department of Physics, 
             University of Maryland,
             College Park, MD 20742, USA}
\date{August 27, 2005}

\begin{abstract}
The design and implementation of large sets of spatially-extended,
gauge-invariant operators for use in determining the spectrum of baryons
in lattice QCD computations are described.  Group-theoretical projections
onto the irreducible representations of the symmetry group of a cubic
spatial lattice are used in all isospin channels.  The operators are 
constructed to maximize overlaps with the low-lying states of interest,
while minimizing the number of sources needed in computing the required
quark propagators. Issues related to the identification of the spin
quantum numbers of the states in the continuum limit are addressed.
\end{abstract}
\maketitle

\section{Introduction}
Spectroscopy is a powerful tool for uncovering the important degrees of
freedom of a physical system and the interaction forces between them. The
spectrum of quantum chromodynamics (QCD) is indeed very rich: conventional
baryons (nucleons, $\Delta, \Lambda, \Xi,\Omega, \dots$) and mesons
$(\pi, K, \rho, \phi,\dots)$  have been known for nearly half a century,
but other higher-lying {\em exotic} states, such as glueballs, four-quark
states, and so-called hybrid mesons and hybrid baryons bound by an excited
gluon field, have proved more elusive, mainly because our theoretical
understanding of such states is insufficient, making their identification
problematical.

Recent discoveries of several new hadronic resonances have generated much
excitement in the field of hadron spectroscopy.  The E852 collaboration\cite{E852}
reported a signal for a $1^{-+}$ hybrid meson decaying into $\rho\pi$ with a
mass near 1.6 GeV, though the significance of this result has been
questioned\cite{dzierba}.  Another exotic $1^{-+}$ candidate at 1.4 GeV has been
tentatively identified in the $\eta\pi$ channel by E852\cite{E852light},
VES\cite{VES}, KEK\cite{KEK}, and Crystal Barrel\cite{CBarrel}.  The first
observation of a doubly-charmed baryon has been reported\cite{mattson}, and
evidence for the possible existence of a strangeness $S=1$ pentaquark state
has emerged at SPring-8, JLab, and elsewhere\cite{CLAS,LEPS,SAPHIR,DIANA}.
Interest in excited baryon resonances has also been raised by experiments
dedicated to mapping out the $N^\ast$ spectrum in Hall B at the Thomas
Jefferson National Accelerator Facility (JLab), and the search for hybrid
mesons and glueballs is intensifying due to the Hall D initiative at JLab
and experiments at CLEO-c.

Unfortunately, our understanding of many conventional and excited hadron 
resonances is vestigial and comes only from QCD-inspired phenomenological 
models, such as the bag model, the nonrelativistic quark model, and
quark-diquark models, or from approaches, such as QCD sum rules and methods
based on Schwinger-Dyson equations, which use approximations whose
justifications are unclear.  There are a growing number of resonances which
cannot be easily accommodated within quark models.  States bound by an 
excited gluon field, such as hybrid mesons and baryons, are still poorly
understood.  The natures of the Roper resonance and the $\Lambda(1405)$
remain controversial.  Experiment shows that the first excited 
positive-parity spin-$1/2$ baryon lies below the lowest-lying 
negative-parity spin-$1/2$ resonance, a fact which is difficult to 
reconcile in quark models.  Given the current surge in experimental activity,
the need for an understanding of such states from QCD itself has never been
greater.  

This need has motivated us to undertake a comprehensive {\em ab initio} study
of the hadron spectrum in QCD.  Presently, Monte Carlo estimates of QCD path
integrals defined on a space-time lattice offer the best way to make progress
in this regard, so this is the calculational approach we have adopted.
The Monte Carlo approach has been used to investigate hadrons throughout
the past two decades, but the number of states studied to date has been 
somewhat limited.  Also, prior works have not strived to identify the 
continuum spin $J$ of the states studied, simply assuming that the lowest
allowed $J$ would be the lowest-lying state.  The novel features of our approach
are its comprehensiveness and its use of techniques to identify spin.  Given the
vast amount of experimental data being generated at accelerator facilities, such
as Jefferson Lab, there is an urgency to investigate the spectrum (masses, widths,
transition rates, form factors, and so on) as completely as possible.  The
number of hadron eigenstates which can be reliably extracted in Monte Carlo
computations is not currently known, so our undertaking will be partly an
exploration of the limits of possibility.  Another aim is to discover whether
a very large number of interpolating operators are needed to extract the 
low-lying spectrum, or whether a handful of carefully chosen ones is sufficient,
and this work outlines a systematic means of finding such operators.

Our first goal is to calculate the masses of as many low-lying hadron resonances
as possible in QCD using Monte Carlo techniques.   The masses and widths
of resonances (unstable hadrons) cannot be calculated directly in finite-volume 
Monte Carlo computations, but must be deduced from the discrete spectrum of
finite-volume stationary states for a range of box 
sizes\cite{dewitt,wiese88,luscher91B,rum95}.  The rigorous application of
such techniques to obtain the resonance parameters to high accuracy would
require vast computational resources, but our goal here is merely to obtain
a first exploratory scan of the spectrum of QCD, not to pin down each mass to
very high precision.  Hence, simply obtaining the finite-volume spectrum for a
few judiciously-chosen volumes should suffice for ferreting out the hadron
resonances from the less interesting scattering states and may even give
qualitative information about preferred decay modes.

To compute the finite-volume stationary state energies, the temporal correlations
$C_{ij}(t)=\langle 0\vert\,T\, O_i(t)\,\overline{O}_j(0)\,\vert 0\rangle$,
where $T$ denotes time-ordering, of 
a set of operators $\overline{O}_j(0)$ which create the states of interest
at an initial time $t=0$ with a corresponding set of operators $O_i(t)$ which
annihilate the states of interest at a later time $t$ must be determined.  
The correlation functions $C_{ij}(t)$ can 
be expressed in terms of path integrals over the quark and gluon fields, and 
when formulated on a Euclidean space-time lattice, such path integrals can be 
estimated using the Monte Carlo method with Markov-chain importance sampling.
Incorporating the quark-field effects into the Monte Carlo updating
for realistically light quark masses remains a challenge, but there is
steady progress with improving algorithms and increasing computational
power.

The procedure for extracting the lowest stationary-state energies 
$E_0,E_1,E_2,\dots$ from the hermitian matrix of correlation functions 
$C_{ij}(t)$ is well known\cite{cmichael, luscherwolff}.
Let $\lambda_n(t,t_0)$ denote the eigenvalues of the hermitian matrix 
$C(t_0)^{-1/2}\,C(t)\,C(t_0)^{-1/2}$, where $t_0$ is some fixed reference time
(typically small) and the eigenvalues, also known as the {\em principal} 
correlation functions, are ordered such that $\lambda_0\geq\lambda_1\geq\cdots$ 
as $t$ becomes large.  Then one can show that
\begin{eqnarray}
 \lim_{t\rightarrow\infty}\lambda_n(t,t_0) &=& e^{-E_n (t-t_0)}\Bigl(
  1 + O(e^{-\Delta_n (t-t_0)})\Bigr),\\
  \Delta_n &=& \min_{k\neq n}\vert E_k-E_n\vert.
\end{eqnarray}
Determinations of the principal correlators $\lambda_n(t,t_0)$ for sufficiently 
large temporal separations $t$ yield the desired energies $E_n$.

The above equations illustrate the difficulties which must be faced in order to 
extract the stationary state energies $E_0,E_1,\dots$ from the temporal 
correlations of the hadronic operators.  In each principal correlator, there are
contaminating contributions from all other states which can be created and 
annihilated by the operators used.  In order to reliably extract the single 
decaying exponential of interest, the obscuring contributions from all of these
other states must somehow be suppressed.

There are two crucial ingredients in reducing the unwanted contributions in 
$\lambda_n(t,t_0)$.  The first is to use sufficiently large values of $t$.  
However, there are often practical considerations which limit how large $t$ 
can be, and the statistical uncertainties in the Monte Carlo estimates of the
correlation functions generally increase with $t$.  The second, and more 
important, consideration in suppressing the contamination in $\lambda_n(t,t_0)$
is to use cleverly-devised operators which couple minimally to the unwanted
states.

Successfully extracting the spectrum of QCD in our Monte Carlo computations
will hinge crucially on using carefully designed hadronic operators.  Excited
meson and baryon resonances are expected to be large objects, so
the use of spatially-extended operators is important.  Since our calculations
will be carried out on hypercubic space-time lattices, the energies of
states in all irreducible representations (irreps) of the $O_h$ cubic point group
must be determined in order to identify the continuum-limit spin $J$ of 
each physical state.  Because determining the mass of a particular resonance
requires determining the energies of all lower-lying stationary states, 
including scattering states, the set of operators we use must include not 
only meson and baryon operators, but also multi-hadron operators.  Another
very important fact to keep in mind is the computational cost of
evaluating quark propagators, especially for light quark masses.  Hence,
our operators must be devised with an eye towards minimizing the
number of sources needed to calculate the required quark propagators.

Designing the hadronic operators is an important first step in our
comprehensive study of the mass spectrum in lattice QCD.  From the above
considerations, the guiding principles in devising our operators are
maximizing overlaps with the states of interest while minimizing the number
of quark-propagator sources.  Although operators for baryons, mesons, and 
their scattering states will be needed, we restrict our attention to the
construction of the three-quark baryon operators in this first paper.  
Due to the complexity of these calculations and the importance of providing checks
on our final results, we have been pursuing two different approaches to 
constructing the baryon interpolating field operators.  Only one approach is
described here; an alternative approach based on Clebsch-Gordan techniques
will appear elsewhere\cite{maryland}.
Meson and multi-hadron operators will be detailed in subsequent works.
Furthermore, this paper deals only with issues related to the construction
and utilization of these operators.  Results from Monte Carlo calculations
using these operators will be presented in later publications.

This paper is organized as follows.  An overview of our approach to
constructing the hadronic operators is first outlined in 
Sec.~\ref{sec:overview}.  Our operators are assemblages of basic building
blocks, which are described in Sec.~\ref{sec:buildblocks}, along with
the conventions we use for the Dirac $\gamma$-matrices and in Wick rotating
into Euclidean space-time.  We use quark fields which are Dirac spinors,
so our hadron operators apply to Monte Carlo calculations involving
Wilson, domain-wall, and overlap fermion actions, but not to computations
involving staggered fermions.  The basic building blocks are then combined
into gauge-invariant three-quark operators (referred to as elemental 
operators) having appropriate flavor structure in Sec.~\ref{sec:elemental}.
Projections onto the rotation-reflection symmetry sectors produce the
final operators in Sec.~\ref{sec:projections}.  Issues related to the 
use of these projections in constructing the baryon propagators are
discussed in Sec.~\ref{sec:baryonprop}.  Concluding remarks and plans for
future work are outlined in Sec.~\ref{sec:conclude}.

\section{Overview of operator construction}
\label{sec:overview}

Devising relativistic hadronic operators in continuous space-time usually
involves combining Dirac spinors to form Lorentz scalars, pseudoscalars, 
vectors, axial-vectors, and so on, using the Dirac $\gamma$-matrices
and the charge conjugation matrix $C$ satisfying
$ C\ \gamma_\mu\ C^{-1}=-\gamma_\mu^T$.
It has been common practice in lattice QCD simulations to use
hadron operators built in a similar fashion through simply discretizing
their continuum analogs.  However, such an
approach becomes very cumbersome when constructing higher spin states
or complicated extended operators.  Also, the above operators generally
couple to states belonging to different $J^P$ (spin-parity) symmetry sectors.
Since the hypercubic lattice breaks Lorentz covariance, there
is really no reason to construct operators according to Lorentz
symmetries.  Finally, we wish to extract a large portion of the low-lying
spectrum, which means that large sets of operators will be needed to 
compute complete correlation matrices.  Hence, the usual approach which
mimics that used in continuous space-time is not feasible for our purposes.

Instead, we advocate an approach which more directly combines the physical
characteristics of baryons with the symmetries of the lattice
regularization of QCD.  Recall that baryon states are characterized by 
their total momentum $\pvec$, their total (half-integral) spin $J$,
a projection $\lambda$ of this spin onto some axis (the $z$-axis or the
momentum, say), their parity $P=\pm 1$, and their quark flavor content.  
The masses of the light $u$ and $d$ quarks are very nearly equal, and therefore,
we work in the approximation that $m_u=m_d$. In this approximation, the theory
has an exact isotopic spin symmetry, and states carry two more labels, 
total isospin $I$ and its projection $I_3$ onto a given axis.  The other
flavor quantum numbers which label the states are strangeness $S$, charm $C$,
and bottomness $B$ (we do not consider the top quark).  In our calculations,
isospin remains an exact symmetry since we neglect electromagnetic interactions.  

Since our simulations will be performed using a hypercubic space-time lattice,
our operators should be classified according to the symmetries of the lattice, 
rather than the full rotational symmetries of continuous space-time. Of course,
we expect to recover the symmetries of the space-time continuum as the lattice
spacing is made small.  Since we are interested only in determining the masses
of the baryon states, we restrict our attention to representations corresponding
to zero total three-momentum $\pvec=\bm{0}$.  Hence, our operators must be
invariant under all allowed spatial translations and we require that they
transform under spatial rotation-reflection symmetry operations
according to the irreducible representations of the octahedral point
group $O_h$. These irreducible representations are the lattice analogs of the
continuum $J^P$ labels, and the row of the representation is the
analog of the spin projection $\lambda$.  Thus, our (annihilation)
operators can be written
\begin{equation}
  B^{\Lambda\lambda F}_i(t)=\sum_{\xvec} B^{\Lambda\lambda F}_i(\xvec,t),
\label{eq:irrepops}
\end{equation}
where $\Lambda$ indicates the irreducible representation of $O_h$, $\lambda$
is the row of the $\Lambda$ representation, $F$ denotes all of the
quantum numbers associated with the flavor content of the operator,
and $i$ labels the different operators in the $\Lambda\lambda F$ symmetry
sector.  Under a symmetry operation $R$, these operators transform 
according to
\begin{eqnarray}
 U_R\ B_i^{\Lambda\lambda F}(t)\ U_R^\dagger
 &=& \sum_\mu B^{\Lambda\mu F}_i(t)\ \Gamma^{(\Lambda)}_{\mu\lambda}(R)^\ast,\\
 U_R\ \overline{B}_i^{\Lambda\lambda F}(t)\ U_R^\dagger
 &=& \sum_\mu \overline{B}^{\Lambda\mu F}_i(t)
 \ \Gamma^{(\Lambda)}_{\mu\lambda}(R),
\end{eqnarray}
where $U_R$ denotes the quantum operator which effects the symmetry
operation corresponding to group element $R$ (not to be confused with the
gauge link variables), and $\Gamma^{(\Lambda)}_{\mu\lambda}(R)$ are the elements
of the $\Lambda$ representation matrix corresponding to group element $R$. 
For baryons, we are only interested in states corresponding to half-integral
spin $J$ in the continuum limit, so we can restrict our attention to the six
spinorial representations of $O_h$.  There are four two-dimensional 
irreducible representations $G_{1g}, G_{1u}, G_{2g}$, and $G_{2u}$ (adopting a
Mulliken-like naming convention), and two four-dimensional representations 
$H_g$ and $H_u$.  These representations will be discussed in greater detail
later.

Our general approach to constructing the $B^{\Lambda\lambda F}_i(t)$ operators
is to (1) first identify appropriate basic ``building blocks'' to use in
constructing all baryon operators, (2) devise simple {\em elemental}
operators containing the appropriate flavor and color structure, then (3)
apply appropriate group-theoretical projection operators to find linear
combinations of the elemental operators with the desired transformation
properties under the symmetry group of a spatial cubic lattice. Let 
$B^F_i(t)=\sum_{\bm{x}}B^F_i(\bm{x},t)$ denote a gauge-invariant elemental 
operator with the appropriate quark flavor content and which is invariant
under allowed spatial translations, then an operator which transforms according
to the row $\lambda$ of the $\Lambda$ irreducible representation is obtained
using
\begin{equation}
  {\cal B}_{i}^{\Lambda\lambda F}(t)
 = \frac{d_\Lambda}{g_{O_h^D}}\sum_{R\in O_h^D} 
  \Gamma^{(\Lambda)}_{\lambda\lambda}(R)\ U_R\ B^F_i(t)\ U_R^\dagger,
\label{eq:project}
\end{equation}
where $O_h^D$ is the double group of $O_h$, $R$ denotes an element of $O_h^D$,
$g_{O_h^D}$ is the number of elements in $O_h^D$, and $d_\Lambda$ is the
dimension of the $\Lambda$ irreducible representation.  Projections onto
the double-valued irreps of $O_h$ require using the double group
$O_h^D$ in Eq.~(\ref{eq:project}). 
Given $M_B$ elemental $B^F_i$ operators, many of the projections in 
Eq.~(\ref{eq:project}) vanish or lead to linearly-dependent operators,
so one must then choose suitable linear combinations of the projected operators
to obtain a final set of independent baryon operators.
Thus, in each symmetry channel, one ends up with a set of $r$ operators 
given in terms of a linear superposition of the $M_B$ elemental operators:
\begin{equation}
  B_i^{\Lambda\lambda F}(t)
 = \sum_{j=1}^{M_B}\ c^{\Lambda\lambda F}_{ij}\ B^F_j(t),\qquad i=1\dots r.
\label{eq:projectcoefs}
\end{equation}
Note that the expansion coefficients in the $\overline{B}^{\Lambda\lambda F}_i$
operators are the complex conjugates of those in $B^{\Lambda\lambda F}_i$:
\begin{equation}
  \overline{B}_i^{\Lambda\lambda F}(t)
 = \sum_{j=1}^{M_B}\ c^{\Lambda\lambda F\ast}_{ij}\ \overline{B}^F_j(t),
 \qquad i=1\dots r.
\label{eq:projectcoefsbar}
\end{equation}

\section{The basic building blocks}
\label{sec:buildblocks}

The oscillating path integral weight $e^{iS_M}$ in quantum field theory,
where $S_M$ is the action defined in Minkowski space-time and using natural
units $\hbar=c=1$, is unsuitable for
applying the Monte Carlo method to evaluate the correlation functions of the
theory via Feynman path integrals.  A rotation to imaginary time
$t\rightarrow -i\tau$, where $\tau$ is real, leads to path integrals with 
weight $e^{-S}$, where $S$ is the action defined in Euclidean space-time.
If $S$ is real, the path integral weight is real and positive and, hence,
can be interpreted as a probability, allowing the application of Monte
Carlo methods with suitable importance sampling.  The Euclidean action $S$ is
defined such that $S$ is invariant
under all symmetries of Euclidean space-time and all Green's functions of the
theory are identical to the Green's functions of the Minkowski theory,
analytically continued to imaginary time $t\rightarrow -i\tau$.  Although our
simulations employ a path integral quantization of the field theory, a 
canonical quantization viewpoint can be adopted when discussing the quantum
operators.

Our conventions for the continuation from Minkowski space-time with metric 
$g_{\mu\nu}={\rm diag}(1,-1,-1,-1)$ into Euclidean space-time (imaginary time) 
are as follows.  We define the following Euclidean space-time coordinates and
derivatives (a subscript or superscript $M$ indicates a Minkowski space-time 
quantity):
\begin{eqnarray}
 && x^4=x_4=ix^0_M,\quad
    x^j=x_j=x^j_M=-x^M_j, \\
 && \partial^4=\partial_4=-i\partial_0^M,\quad
 \partial^j=\partial_j=-\partial^j_M=\partial_j^M,
\end{eqnarray}
for spatial directions $j=1,2,3$.
The metric in Euclidean space-time is $\delta^{\mu\nu}$ so there is no distinction
between covariant and contravariant indices.   Our Monte Carlo calculations will
be carried out using a hypercubic space-time lattice, and we require that the 
lattice spacings in the three spatial directions are the same, denoted 
by $a_s$; the temporal lattice spacing $a_t$ may differ from $a_s$, allowing us to
exploit the known benefits of anisotropic lattices\cite{colin2}.  
Throughout this paper, we set
$a_s=1$ to simplify the notation.  Four-vectors of unit length pointing along 
the spatial axes of the lattice with a vanishing temporal component will be 
denoted by $\hat{j}$, $\hat{k}$, and so on, for $j,k=\pm 1,\pm 2,\pm 3$.

As usual in lattice gauge theory, the gluon field is introduced using
the parallel transporter $U_\mu(x)$ given by the path-ordered exponential
of the gauge field along each link connecting neighboring sites of the
hypercubic lattice.  We also introduce the Dirac spinor field
$\psi^A_{a\alpha}(x)$ which annihilates a quark and creates an antiquark,
where $A$ refers to the quark flavor, $a$ refers to color, and $\alpha$ is 
the Dirac spin index, and the field $\overline{\psi}^A_{a\alpha}(x)$
which annihilates an antiquark and creates a quark.  Unlike in Minkowski 
space-time, $\psi$ and $\overline{\psi}$ must be treated as independent fields,
so we emphasize that $\overline{\psi}\neq \psi^\dagger\gamma_4$.
This is required in order to simultaneously satisfy Euclidean covariance of
the fields, the canonical anticommutation relations, and the equality of the
Euclidean two-point function with the relativistic Feynman propagator
continued to imaginary times\cite{osterwalder,williams}.  Our Euclidean space-time
Dirac-$\gamma$ matrices are related to their Minkowski counterparts by
\begin{eqnarray}
  &&  \gamma^4= \gamma_4=\gamma^0_M,
     \qquad \gamma_k=\gamma^k=-i\gamma^k_M,\\
      &&  \{\gamma_\mu,\gamma_\nu\}=2\delta_{\mu\nu},
      \qquad \gamma^\dagger_\mu=\gamma_\mu,\\
    &&  \qquad \gamma^5=\gamma_5
         =\gamma_4 \gamma_1\gamma_2\gamma_3 =\gamma^5_M.
\end{eqnarray}
Throughout this paper, we use the standard Dirac-Pauli representation for the
$\gamma$-matrices:
\begin{equation}
    \gamma_k = \left(\begin{array}{cc} 0 &-i\sigma_k
   \\ i\sigma_k& 0\end{array}\right)\!,\ \ \ \gamma_4=
   \left(\begin{array}{cc} I &0 \\0&-I
     \end{array}\right)\!,
\end{equation}
where the Pauli spin matrices are given by
\begin{equation}
 \sigma_1 = \left(\begin{array}{rr}0&1\\1&0\end{array}\right)\!,
 \ \sigma_2 = \left(\begin{array}{rr}0&-i\\i&0\end{array}\right)\!,
 \ \sigma_3 = \left(\begin{array}{rr}1&0\\0&-1\end{array}\right)\!.
\end{equation}

It has long been known that operators constructed out of smeared fields have 
dramatically reduced mixings with the high frequency modes of the theory.  
Thus, our operators are constructed using spatially-smoothed link variables
$\widetilde{U}_j(x)$ and spatially-smeared quark fields $\widetilde{\psi}(x)$.
The spatial links are smeared using either the stout-link procedure
described in Ref.~\cite{stout} or the method introduced in Ref.~\cite{APEsmear}.
Note that only spatial staples are used in the link smoothening; no temporal
staples are used, and the temporal link variables are not smeared.
The smeared quark fields are defined by\cite{quarksmear}
\begin{eqnarray}
\widetilde{\psi}(x) &=& \left(1+\frac{\sigma_s^2}{4n_\sigma}\ \widetilde{\Delta}
 \right)^{n_\sigma}\psi(x),\\
\widetilde{\overline{\psi}}(x) &=& \overline{\psi}(x)
 \left(1+\frac{\sigma_s^2}{4n_\sigma}
 \ \overleftarrow{\widetilde{\Delta}} \right)^{n_\sigma},
\end{eqnarray}
where $\sigma_s$ and $n_\sigma$ are tunable parameters ($n_\sigma$ is 
a positive integer) and the three-dimensional covariant Laplacian
operators are defined in
terms of the smeared link variables $\widetilde{U}_j(x)$ as follows: 
\begin{eqnarray}
 \widetilde{\Delta} O(x) &=& \!\!\!\sum_{k=\pm 1,\pm2,\pm3} \biggl(
  \widetilde{U}_k(x)O(x\!+\!\hat{k})-O(x) \biggr), \\
  \overline{O}(x)\overleftarrow{\widetilde{\Delta}} &=& \!\!\!\sum_{k=\pm 1,\pm2,\pm3}
 \biggl(  \overline{O}(x\!+\!\hat{k})\widetilde{U}_k^\dagger(x)
 -\overline{O}(x) \biggr),
\end{eqnarray}
where $O(x)$ and $\overline{O}(x)$ are operators defined at lattice site
$x$ with appropriate color structure, and noting that
$\widetilde{U}_{-k}(x)=\widetilde{U}_k^\dagger(x\!-\!\hat{k})$.  The smeared
fields $\widetilde{\psi}$ and $\widetilde{\overline{\psi}}$ are Grassmann-valued;
in particular, these fields anticommute in the same way that the original
fields do, and the square of each smeared field vanishes.

Our operators are designed with one eye on capturing the states of interest
and the other eye on facilitating the efficient computation of the hadron
correlation functions.  Since the baryon resonances are expected to be large
objects, the use of extended operators is crucial.  Our choice
of basic building blocks is motivated by the need to incorporate spatial
extensions so as to facilitate the efficient and gauge-invariant 
assembly of many large hadron operators.  To capture orbital structure, the
quarks must be displaced in different directions, and to capture radial 
structure, the quarks must be displaced by different distances.  To maintain
gauge invariance, covariant displacements in terms of the gauge-field parallel
transporters must be used.  To simplify matters, we consider only straight-path
displacements in the six directions along the axes of the spatial cubic lattice:
$j=\pm 1,\pm 2,\pm 3$.
Hence, we shall assemble our baryon (and later, meson) operators
using the following basic building blocks:
\[
  \widetilde{\psi}^A_{a\alpha},\  \widetilde{\overline{\psi}}^A_{a\alpha},
 \ \bigl(\widetilde{D}^{(p)}_j\ \widetilde{\psi}\bigr)^A_{a\alpha},
 \  \bigl(\widetilde{\overline{\psi}}  
   \ \widetilde{D}^{(p)\dagger}_j \bigr)^A_{a\alpha},
  \quad j=\pm1,\pm2,\pm3,
\]
where $A$ is a flavor index, $a$ is a color index, $\alpha$ is a
Dirac spin index, and the $p$-link gauge-covariant displacement
operator in the $j$-th direction is defined by
\begin{equation}
 \widetilde{D}_j^{(p)}(x,x^\prime) =
 \widetilde{U}_j(x)\ \widetilde{U}_j(x\!+\!\hat{j})\dots 
   \widetilde{U}_j(x\!+\!(p\!-\!1)\hat{j})\delta_{x^\prime,x+p\hat{j}},
\end{equation}
for $j=\pm 1,\pm 2,\pm 3$ and $p\geq 1$.  In what follows, we can achieve 
a significant economy in the notation by defining a zero-displacement operator
\begin{equation}
\widetilde{D}_0^{(p)}(x,x^\prime) = \delta_{xx^\prime},
\end{equation}
to indicate no displacement. In this way, our basic building
blocks can be listed more succinctly:
\begin{equation}
\bigl(\widetilde{D}^{(p)}_j\ \widetilde{\psi}\bigr)^A_{a\alpha},
 \  \bigl(\widetilde{\overline{\psi}}  
   \ \widetilde{D}^{(p)\dagger}_j \bigr)^A_{a\alpha},
 \qquad -3\leq j\leq 3.
\label{eq:blocks}
\end{equation}
Sometimes it will be convenient to make the 
flavor quantum number more apparent by writing
\begin{equation}
\widetilde{u}_{a\alpha}(x)\equiv \widetilde{\psi}^u_{a\alpha}(x), \quad
\widetilde{d}_{a\alpha}(x)\equiv \widetilde{\psi}^d_{a\alpha}(x),
\end{equation}
and similarly for the $s,c,b$ quarks.  It is important to remember that the
$\widetilde{u},\widetilde{d},\widetilde{s},$ {\em etc.}\ operators 
so defined refer to smeared quark fields.

\section{Three-quark elemental operators}
\label{sec:elemental}

Having chosen the basic building blocks for our hadron operators listed
in Eq.~(\ref{eq:blocks}), the
next step is to devise {\em elemental} baryon operators $B^F_i(t)$
having the appropriate color and flavor structure.  These operators
are chosen such that they are gauge-invariant and transform irreducibly
under the isotopic spin symmetry.  Explicitly dealing with
$SU(3)$ flavor symmetry is not necessary, as will be discussed below.

Gauge invariance is easily handled.  Given three quarks with
color indices $a,b,c$ associated with the same lattice site $x$, there is
only one way of combining the color indices to arrive at a locally
gauge-invariant object: the use of the antisymmetric Levi-Civita
symbol $\varepsilon_{abc}$.  Similarly, covariantly-displaced quark fields
must also be connected by an $\varepsilon_{abc}$ coupling at a single
lattice site.  To simplify the operator construction, we consider only
combinations of displaced quarks having the same displacement length $p$.
Thus, all of our three-quark baryon operators are linear superpositions
of gauge-invariant terms of the form
\begin{eqnarray}
&\Phi^{ABC}_{\alpha\beta\gamma;\,ijk}(t)
 = \sum_{\bm{x}}\   \varepsilon_{abc} 
  \ \Bigl((\widetilde{D}^{(p)}_i\widetilde{\psi})^A_{a\alpha}(\xvec,t) &\nonumber\\
  &\qquad\quad\times  (\widetilde{D}^{(p)}_j\widetilde{\psi})^B_{b\beta}(\xvec,t)
  \  (\widetilde{D}^{(p)}_k\widetilde{\psi})^C_{c\gamma}(\xvec,t)\Bigr).&
 \label{eq:giform}
\end{eqnarray}
The ``barred'' operators have the form
\begin{eqnarray}
&\overline{\Phi}^{ABC}_{\alpha\beta\gamma;\,ijk}(t)
 = \sum_{\bm{x}}\   \varepsilon_{abc} 
  \ \Bigl((\widetilde{\overline{\psi}}
\widetilde{D}^{(p)\dagger}_k)^C_{c\gamma^\prime}(\xvec,t)
  \gamma^4_{\gamma^\prime\gamma}& \nonumber\\
  &\times 
(\widetilde{\overline{\psi}}\widetilde{D}^{(p)\dagger}_j)^B_{b\beta^\prime}(\xvec,t)
 \gamma^4_{\beta^\prime\beta}
  \  (\widetilde{\overline{\psi}}\widetilde{D}^{(p)\dagger}_i)^A_{a\alpha^\prime}
(\xvec,t)\gamma^4_{\alpha^\prime\alpha}\Bigr).&
\label{eq:barops}
\end{eqnarray}
Note that the barred composite operator $\overline{\Phi}$ is defined 
differently than the barred fermion field operator $\overline{\psi}$ due
to the presence of the $\gamma_4$ spin matrices in Eq.~(\ref{eq:barops}).
Their presence is needed to ensure that the resulting correlation matrices
satisfy the desirable property of hermiticity.  These $\gamma_4$ matrices
do not affect the transformation properties of these operators under proper
spatial rotations and parity, although they do affect the Lorentz boost
properties.

\begin{table}[t]
\caption[captab]{The six types of three-quark 
$\Phi^{ABC}_{\alpha\beta\gamma;\ ijk}$ operators used, where $A,B,C$ indicate
the quark flavors, $1\leq \alpha,\beta,\gamma\leq 4$ are Dirac spin indices, and
$-3\leq i,j,k\leq 3$ are displacement indices. Smeared quark fields are
shown by solid circles, line segments indicate covariant displacements, and each
hollow circle indicates the location of a color $\varepsilon_{abc}$ coupling.  
To simplify matters, all displacements have the same length ($p$ lattice links) in any
given operator.  Remember that a displacement index having a zero value indicates
no displacement.  The singly-displaced operators are meant to mock up a quark-diquark
combination, and the doubly-displaced and triply-displaced operators are chosen since
they may favor $\Delta$-flux and $Y$-flux configurations, respectively.
\label{tab:opforms}}
\begin{ruledtabular}
\begin{tabular}{cl}
 Operator type &  Displacement indices\\ \hline
\raisebox{0mm}{\setlength{\unitlength}{1mm}
\thicklines
\begin{picture}(16,10)
\put(8,6.5){\circle{6}}
\put(7,6){\circle*{2}}
\put(9,6){\circle*{2}}
\put(8,8){\circle*{2}}
\put(0,0){single-site}
\end{picture}}  & \raisebox{3mm}{$i=j=k=0$ }\\ 
\raisebox{0mm}{\setlength{\unitlength}{1mm}
\thicklines
\begin{picture}(23,10)
\put(7,6.2){\circle{5}}
\put(7,5){\circle*{2}}
\put(7,7.3){\circle*{2}}
\put(14,6){\circle*{2}}
\put(9.5,6){\line(1,0){4}}
\put(0,0){singly-displaced}
\end{picture}}  & \raisebox{3mm}{$i=j=0,\ k\neq 0$} \\ 
\raisebox{0mm}{\setlength{\unitlength}{1mm}
\thicklines
\begin{picture}(26,8)
\put(12,5){\circle{3}}
\put(12,5){\circle*{2}}
\put(6,5){\circle*{2}}
\put(18,5){\circle*{2}}
\put(6,5){\line(1,0){4.2}}
\put(18,5){\line(-1,0){4.2}}
\put(-1,0){doubly-displaced-I}
\end{picture}}  & \raisebox{2mm}{$i=0,\ j=-k,\ k\neq 0$} \\ 
\raisebox{0mm}{\setlength{\unitlength}{1mm}
\thicklines
\begin{picture}(20,13)
\put(8,5){\circle{3}}
\put(8,5){\circle*{2}}
\put(8,11){\circle*{2}}
\put(14,5){\circle*{2}}
\put(14,5){\line(-1,0){4.2}}
\put(8,11){\line(0,-1){4.2}}
\put(-5,0){doubly-displaced-L}
\end{picture}}   & \raisebox{4mm}{$i=0,\ \vert j\vert\neq \vert k\vert,
  \ jk\neq 0$}\\ 
\raisebox{0mm}{\setlength{\unitlength}{1mm}
\thicklines
\begin{picture}(20,12)
\put(10,10){\circle{2}}
\put(4,10){\circle*{2}}
\put(16,10){\circle*{2}}
\put(10,4){\circle*{2}}
\put(4,10){\line(1,0){5}}
\put(16,10){\line(-1,0){5}}
\put(10,4){\line(0,1){5}}
\put(-5,0){triply-displaced-T}
\end{picture}}   & \raisebox{4mm}{$i=-j,\ \vert j\vert \neq\vert k\vert,
 \ jk\neq 0$} \\ 
\raisebox{0mm}{\setlength{\unitlength}{1mm}
\thicklines
\begin{picture}(20,12)
\put(10,10){\circle{2}}
\put(6,6){\circle*{2}}
\put(16,10){\circle*{2}}
\put(10,4){\circle*{2}}
\put(6,6){\line(1,1){3.6}}
\put(16,10){\line(-1,0){5}}
\put(10,4){\line(0,1){5}}
\put(-5,0){triply-displaced-O}
\end{picture}}   & \raisebox{4mm}{$\vert i\vert \neq \vert j\vert \neq
  \vert k\vert,\ ijk\neq 0$} 
\end{tabular}
\end{ruledtabular}
\end{table}

The simplest way to combine the basic building blocks previously
described is to combine three quark fields at a single lattice site,
corresponding to $i=j=k=0$ in Eq.~(\ref{eq:giform}).  We refer to these
operators as single-site operators.  Next, one of the three quarks can be
displaced; such singly-displaced operators correspond to $i=j=0, k\neq 0$
in Eq.~(\ref{eq:giform}).  In these operators, the two quarks which are
not displaced may be viewed as forming a localized diquark, so such operators
may be important if baryon formation is dominated by a quark-diquark mechanism.  

Two of the quarks can be displaced; they can be displaced either
in opposite directions \mbox{(doubly-displaced-I)}, or in orthogonal directions
\mbox{(doubly-displaced-L)}.  In such operators, one can in general choose 
different lengths for the two displacements, but to simplify matters, we
restrict our attention to the case in which both displacements have the same
length $p$.  Since the displacement of two quark fields essentially results
in an object in which all three quarks are at different lattice sites, one
may be tempted to exclude these operators in favor of triply-displaced 
operators.  However, the relative importance of so-called $Y$-flux and 
$\Delta$-flux formation of the gluon field in three-quark systems (the $\Delta$
is actually a quantum superposition of $V$-flux forms) is a much-discussed issue
(see, for example, Refs.~\cite{sommer,bali,forcrand1,forcrand2,forcrand3,
suganuma1,suganuma2}).
Hence, it is important to include some operators with significant overlap with a 
$\Delta$-flux configuration, as well as operators with strong mixing with
a $Y$-flux configuration.  The doubly-displaced operators above are
suitable for $\Delta$-flux formation.

Lastly, all three quarks can be displaced (again, restricting attention
to the case of equal distances).  If two of
the quarks are displaced in opposite directions, this produces a coplanar
T-shape \mbox{(triply-displaced-T)}; alternatively, the three quarks can be 
displaced in mutually orthogonal directions \mbox{(triply-displaced-O)}.
These operators are suitable for producing $Y$-flux configurations.

These operators, summarized and illustrated in Table~\ref{tab:opforms},
allow a large number of baryon operators to be constructed using a relatively small
number of quark propagator sources. For a given reference source site $x$,
quark propagators must be evaluated using only a handful of different
sources: for each quark mass value, we need a local source and displaced
sources in at most each of the six directions.  However, rotational
invariance of the baryon correlation functions can be exploited to
reduce the number of source displacement directions.  For singly-displaced
operators at the source, a simultaneous rotation of the source and sink
can always be used to align the source displacement along the $+z$ direction,
say.  For doubly-displaced-I sources, a rotation can always be done to
align one displacement along the $+z$ direction and the other along the
$-z$ direction.  Doubly-displaced-L sources can be rotated so the source
displacements are along the $+y$ and $+z$ directions, triply-displaced-T
sources can be rotated to align the displacements along the $+y,+z,$ and
$-z$ directions, and triply-displaced-O sources can always be rotated so the
displacements are along the  $+x,+y,+z$ directions.  In total, only source
displacements along the $+x,+y,+z,-z$ directions (four directions) are
required.  Hence, the number of conjugate gradient inversions needed is
\begin{equation}
  N_{\rm inversions}=N_c\ N_{sp}\ N_\kappa (1+4N_p), 
\end{equation}
where $N_c=3$ is the number of colors, $N_{sp}=4$ is the number of
Dirac spin components, $N_p$ is the number of displacement lengths $p$, and
$N_\kappa$ is the number of quark masses to be used.

\begin{table}[t]
\caption[captab]{Elemental baryonic operators which annihilate certain states
 of definite isospin $I$, maximal $I_3=I$, and strangeness $S$, in terms 
 of the gauge-invariant three-quark operators defined in 
 Eq.~(\protect\ref{eq:giform}).
\label{tab:flavors}}
\begin{ruledtabular}
\begin{tabular}{ccrc}
 Baryon & $I=I_3$ & $S$ & Annihilation operators \\ \hline
 $\Delta^{++}$ & $\frac{3}{2}$ & $0$  & $\Phi^{uuu}_{\alpha\beta\gamma;\, ijk}$\\
 $\Sigma^+$ & $1$ & $-1$ & $\Phi^{uus}_{\alpha\beta\gamma;\,ijk}$\\
 $N^+$ &  $\frac{1}{2}$ &  $0$ &
 $\Phi^{uud}_{\alpha\beta\gamma;\,ijk}-\Phi^{duu}_{\alpha\beta\gamma;\,ijk}$ \\
 $\Xi^0$ & $\frac{1}{2}$ & $-2$ & $\Phi^{ssu}_{\alpha\beta\gamma;\,ijk}$\\
 $\Lambda^0$ & $0$ & $-1$ & 
 $\Phi^{uds}_{\alpha\beta\gamma;\,ijk}-\Phi^{dus}_{\alpha\beta\gamma;\,ijk}$ \\
$\Omega^-$ & $0$ & $-3$ & $\Phi^{sss}_{\alpha\beta\gamma;\, ijk}$
\end{tabular}
\end{ruledtabular}
\end{table}

\begin{table}[t]
\caption[captab]{The $\Delta^{++}$ elemental operators contain three 
 $u$ quarks and are given by $\Phi^{uuu}_{\alpha\beta\gamma;\,ijk}$ as 
 defined in Eq.~(\protect\ref{eq:giform}) with spin components 
 $1\leq\alpha,\beta,\gamma\leq 4$ and displacement indices 
 $-3\leq i,j,k\leq 3$.   The linearly independent operators
 we chose are described in the second column, and the numbers of independent
 operators of each type are listed in the third column. Replacing each
 $u$ quark by an $s$ quark in the operators below yields $\Omega^-$ 
 elemental operators.
\label{tab:deltaelemental}}
\begin{ruledtabular}
\begin{tabular}{lcr}
 Operator type & Restrictions & \hspace{-8mm}Multiplicity \\ \hline
  single-site &
 $(i\!=\!j\!=\!k\!=\!0,\,\alpha\leq\beta\leq\gamma)$ & 20 \\
  singly-displaced &
 $(i\!=\!j\!=\!0,k\!\neq\!0,\,\alpha\leq\beta)$ & 240 \\
 doubly-displaced-I &
   $(i\!=\!0,\,j=-k,\ k>0)$ &  192 \\
 doubly-displaced-L &
  $\ (i\!=\!0,\,jk\neq 0;\ \vert j\vert\!<\!\vert k\vert)$ & 768\\
  triply-displaced-T &
 $\ (i=-j,\ j>0, \ j\neq \vert k\vert,\ k\neq 0)$ & 768 \\ 
 triply-displaced-O &
 $\ (ijk\neq 0,
    \ \vert i\vert\!\! < \!\!\vert j\vert \!\!<\!\! \vert k\vert)$ & 512 
\end{tabular}
\end{ruledtabular}
\end{table}

\begin{table}[b]
\caption[captab]{The $\Sigma^+$ elemental operators 
 are given by $\Phi^{uus}_{\alpha\beta\gamma;\,ijk}$ as defined in
 Eq.~(\protect\ref{eq:giform}).  The linearly independent operators
 we chose are described in the second column, and the numbers of independent
 operators of each type are listed in the third column.
  Interchanging the $u$ and $s$ quarks 
 yields the $\Xi^0$ elemental operators.
\label{tab:sigmaelemental}}
\begin{ruledtabular}
\begin{tabular}{lcr}
 Operator type & Restrictions & \hspace{-10mm}Multiplicity \\ \hline
 single-site&  $(i\!=\!j\!=\!k\!=\!0,\,\alpha\leq\beta)$ & 40\\
 singly-displaced & $(i\!=\!j\!=\!0,\,k\neq0,\,\alpha\leq\beta)$ & 240 \\
 singly-displaced & $(i\!=\!k\!=\!0,\,j\neq 0)$ & 384 \\
 doubly-displaced-I & $(i\!=\!0,\,j\!=\!-k,\,k\neq0)$ & 384\\
 doubly-displaced-I & $(i\!=\!-j,\,j>0,\,k\!=\!0)$ & 192\\
 doubly-displaced-L & $(i\!=\!0,\, jk\neq0,\,\vert j\vert\neq\vert k\vert)$ & 1536 \\
 doubly-displaced-L & $(ij\neq0,\,\vert i\vert<\vert j\vert,\,k\!=\!0)$ & 768\\ 
 triply-displaced-T & $(i\!=\!-j,\,j>0,\,j\neq\vert k\vert,\,k\neq 0)$ & 768 \\
 triply-displaced-T & $(ij\neq 0,\,\vert i\vert\neq\vert j\vert,\, k\!=\!-i)$ & 1536\\
 triply-displaced-O & $(ijk\!\neq\! 0, \vert i\vert\!\! < \!\!\vert j\vert,
    \,\vert i\vert\!\!\neq\!\!\vert k\vert,
    \,\vert j\vert\!\!\neq\!\!\vert k\vert)$ & 1536
\end{tabular}
\end{ruledtabular}
\end{table}

\begin{table}[t]
\caption[captab]{The $N^+$ elemental destruction operators are given by 
 $\Phi^{uud}_{\alpha\beta\gamma;\,ijk}-\Phi^{duu}_{\alpha\beta\gamma;\,ijk}$
 (see Eq.~(\protect\ref{eq:giform})).  The linearly independent 
 operators we chose are described in the second column, and the numbers of
 independent operators of each type are listed in the third column. 
\label{tab:nucleonelemental}}
\begin{ruledtabular}
\begin{tabular}{lcr}
 Operator type & Restrictions & \hspace{-10mm}Multiplicity \\ \hline
  single-site &
 $(i\!=\!j\!=\!k\!=\!0,\,\alpha\geq\beta,\ \alpha >\gamma)$ & 20 \\
  singly-displaced &
 $(i\!=\!j\!=\!0,k\!\neq\!0)$ & 384 \\
 doubly-displaced-I &
   $(i\!=\!0,\,j=-k,\ k\neq 0)$ &  384 \\
 doubly-displaced-L &
  $\ (i\!=\!0,\,jk\neq 0;\ \vert j\vert\!\neq \!\vert k\vert)$ & 1536\\
  triply-displaced-T &
 $\ (i=-j,\ jk\!\neq\!0,\,\vert j\vert\neq\vert k\vert)$ & 1536 \\ 
 triply-displaced-O &
 $\ (ijk\neq 0,\,\vert i\vert\!\! < \!\!\vert j\vert,
    \,\vert i\vert\!\! < \!\!\vert k\vert,
    \,\vert j\vert\neq\vert k\vert)$ & 1024
\end{tabular}
\end{ruledtabular}
\end{table}

\begin{table}[b]
\caption[captab]{The $\Lambda^0$ elemental destruction operators are given by 
 $\Phi^{uds}_{\alpha\beta\gamma;\,ijk}-\Phi^{dus}_{\alpha\beta\gamma;\,ijk}$
 (see Eq.~(\protect\ref{eq:giform})).  The linearly independent 
 operators we chose are described in the second column, and the numbers of
 independent operators of each type are listed in the third column. 
\label{tab:lambdaelemental}}
\begin{ruledtabular}
\begin{tabular}{lcr}
 Operator type & Restrictions & \hspace{-10mm}Multiplicity \\ \hline
 single-site&  $(i\!=\!j\!=\!k\!=\!0,\,\alpha <\beta)$ & 24\\
 singly-displaced & $(i\!=\!j\!=\!0,\,k\neq0,\,\alpha <\beta)$ & 144 \\
 singly-displaced & $(i\!=\!k\!=\!0,\,j\neq 0)$ & 384 \\
 doubly-displaced-I & $(i\!=\!0,\,j\!=\!-k,\,k\neq0)$ & 384\\
 doubly-displaced-I & $(i\!=\!-j,\,j>0,\,k\!=\!0)$ & 192\\
 doubly-displaced-L & $(i\!=\!0,\, jk\neq0,\,\vert j\vert\neq\vert k\vert)$ & 1536 \\
 doubly-displaced-L & $(ij\neq0,\,\vert i\vert<\vert j\vert,\,k\!=\!0)$ & 768\\ 
 triply-displaced-T & $(i\!=\!-j,\,j>0,\,j\neq\vert k\vert,\,k\neq 0)$ & 768 \\
 triply-displaced-T & $(ij\neq 0,\,\vert i\vert\neq\vert j\vert,\, k\!=\!-i)$ & 1536\\
 triply-displaced-O & $(ijk\!\neq\! 0, \vert i\vert\!\! < \!\!\vert j\vert,
    \,\vert i\vert\!\!\neq\!\!\vert k\vert,
    \,\vert j\vert\!\!\neq\!\!\vert k\vert)$ & 1536
\end{tabular}
\end{ruledtabular}
\end{table}

Incorporating the isospin symmetry is also straightforwardly done.
Let $\tau_1,\tau_2,\tau_3$ denote the three hermitian generators of the isospin
symmetry satisfying the commutation relations
$\lbrack \tau_i,\,\tau_j\rbrack=i\varepsilon_{ijk}\tau_k$.
A set of quantum operators $\overline{O}^{(I)}_{I_3}$ transforms
under isospin according to the irreducible representation $I$ if
\begin{eqnarray}
 \lbrack\tau_3,\ \overline{O}^{(I)}_{I_3}\rbrack &=& I_3\ \overline{O}^{(I)}_{I_3},\\
 \lbrack\tau_+,\ \overline{O}^{(I)}_{I_3}\rbrack 
    &=& \sqrt{(I-I_3)(I+I_3+1)}\ \overline{O}^{(I)}_{I_3+1},\\
 \lbrack\tau_-,\ \overline{O}^{(I)}_{I_3}\rbrack 
   &=& \sqrt{(I+I_3)(I-I_3+1)}\ \overline{O}^{(I)}_{I_3-1},
\end{eqnarray}
where $\tau_\pm=\tau_1\pm i\tau_2$.  It follows from these relations
that
\begin{eqnarray}
 &&\lbrack \tau_3,\ \lbrack \tau_3,\ \overline{O}^{(I)}_{I_3}\rbrack\rbrack
+\textstyle\frac{1}{2} \lbrack \tau_+,\ \lbrack \tau_-,
 \ \overline{O}^{(I)}_{I_3}\rbrack\rbrack \nonumber\\
&&\qquad+\textstyle\frac{1}{2} \lbrack \tau_-,\ \lbrack \tau_+
,\ \overline{O}^{(I)}_{I_3}\rbrack\rbrack
 = I(I+1)\ \overline{O}^{(I)}_{I_3}. \label{eq:isospin1}
\end{eqnarray}
Baryonic operators of definite isospin $I$ and $I_3$ are easily
constructed using the above relations and the following commutation
rules involving the isospin generators and the barred
$\overline{u},\overline{d},\overline{s}$ quark field
operators (suppressing all indices except flavor):
\begin{equation}
\begin{array}{rcl@{\hspace{3mm}}rcl@{\hspace{3mm}}rcl}
\lbrack\tau_3,\, \overline{u}\rbrack&=&\frac{1}{2}\overline{u}, 
  & \lbrack\tau_3,\, \overline{d}\rbrack&=&-\frac{1}{2}\overline{d}, 
  & \lbrack\tau_3,\, \overline{s}\rbrack&=&0,\\
\lbrack\tau_+,\, \overline{u}\rbrack&=&0,
 &   \lbrack\tau_+,\, \overline{d}\rbrack&=&\overline{u}, 
   &             \lbrack\tau_+,\, \overline{s}\rbrack&=&0,\\
\lbrack\tau_-,\, \overline{u}\rbrack&=&\overline{d},
 &   \lbrack\tau_-,\, \overline{d}\rbrack&=&0, 
 &             \lbrack\tau_-,\, \overline{s}\rbrack&=&0,
\end{array}
\end{equation}
and for the unbarred field operators,
\begin{equation}
\begin{array}{rcl@{\hspace{3mm}}rcl@{\hspace{3mm}}rcl}
\lbrack\tau_3,\ u\rbrack&=&-\frac{1}{2}u, 
  & \lbrack\tau_3,\ d\rbrack&=&\frac{1}{2}d, 
  & \lbrack\tau_3,\ s\rbrack&=&0,\\
\lbrack\tau_-,\ u\rbrack&=&0, &   \lbrack\tau_-,\ d\rbrack&=&-u, 
   &             \lbrack\tau_-,\ s\rbrack&=&0,\\
\lbrack\tau_+,\ u\rbrack&=&-d, &   \lbrack\tau_+,\ d\rbrack&=&0, 
 &             \lbrack\tau_+,\ s\rbrack&=&0.
\end{array}
\end{equation}

Due to the isospin symmetry of QCD in the $m_u=m_d$ approximation,
the particle masses do not depend on $I_3$, so we limit
our attention to only one value of $I_3$ in each isospin $I$,
strangeness $S$ sector, choosing the maximal $I_3=I$ value.
Using the above isospin relations, we first write down all
possible flavor combinations appropriate for each isospin channel
for the three-quark elemental operators from Table~\ref{tab:opforms}.
These are listed in Table~\ref{tab:flavors}.  For each flavor sector
and quark-displacement type, we then determine a maximal set of 
linearly independent operators, giving us the final set of elemental
operators we use.  A symbolic computer program capable of manipulating
Grassmann fields was written using {\sc Maple 9.5} and utilized to identify
linearly independent operators.  The independent elemental operators
we chose in the different flavor sections are described in 
Tables~\ref{tab:deltaelemental}, \ref{tab:sigmaelemental},
\ref{tab:nucleonelemental}, and \ref{tab:lambdaelemental}.

Due to an approximate $SU(3)$ $uds$-flavor symmetry, quark flavor combinations
in baryon operators are usually chosen according to the irreducible
representations of $SU(3)$ flavor.  Such combinations are simply linear
superpositions of the operators presented above.  Since we plan to
obtain Monte Carlo estimates of the complete correlation matrices of
operators including all allowed flavor combinations, the use of linear
superpositions which transform irreducibly under $SU(3)$ flavor is
unnecessary.  Our choice of operators described above is dictated
by computational simplicity and efficiency.

Lastly, note that the baryons $\Lambda_c, \Sigma_c, \Xi_{cc}$,
and $\Omega_{ccc}$ can be studied using the operators presented above
if the $s$ quark is replaced with a $c$ quark.  Similarly, replacing
the $s$ quark by a $b$ quark in the above operators allows us to investigate
the $\Lambda_b, \Sigma_b, \Xi_{bb}$, and $\Omega_{bbb}$ baryons.
Some other baryons, such as $\Omega_{cc}$, can be studied using other suitable
flavor replacements in the above operators, whereas the investigations of 
baryons such as $\Xi_c$ containing $usc$ quarks cannot directly exploit the
above tables, requiring slight modifications. 
\vspace{-4mm}

\section{Projections onto symmetry sectors}
\label{sec:projections}

\begin{table}[t]
\caption[jspin]{Continuum limit spin identification:
   the number $n_\Lambda^J$ of times that the $\Lambda$ irrep of the
   octahedral point group $O_h$ occurs in the
   (reducible) subduction of the $J$ irrep of $SU(2)$.  The numbers
   for $G_{1u},G_{2u},H_u$ are the same as for $G_{1g},G_{2g},H_g$,
    respectively.
\label{tab:spin}}
\begin{ruledtabular}
\begin{tabular}{cccc@{\hspace{2em}}cccc}
  $J$  & $n^J_{G_{1g}}$ & $n^J_{G_{2g}}$ & $n^J_{H_g}$ &
 $J$  & $n^J_{G_{1g}}$ & $n^J_{G_{2g}}$ & $n^J_{H_g}$\\ \hline
  $\frac{1}{2}$  &  $1$ & $0$ & $0$ &$\frac{9}{2} $ &  $1$ & $0$ & $2$ \\
  $\frac{3}{2}$  &  $0$ & $0$ & $1$ &$\frac{11}{2}$ &  $1$ & $1$ & $2$ \\
  $\frac{5}{2} $ &  $0$ & $1$ & $1$ &$\frac{13}{2}$ &  1 & 2 & 2 \\
  $\frac{7}{2} $ &  $1$ & $1$ & $1$ &$\frac{15}{2}$ &  1 & 1 & 3
\end{tabular}
\end{ruledtabular}
\end{table}

The final step in our operator construction is to apply group-theoretical
projections to obtain operators which transform irreducibly under all lattice
rotation and reflection symmetries.  The basic building blocks used to assemble
our baryon operators transform under the allowed spatial rotations and reflections
of the point group $O_h$ according to
\begin{eqnarray}
  &U_R \bigl(\widetilde{D}^{(p)}_j
  \widetilde{\psi}(x)\bigr)^A_{a\alpha}\! U_R^\dagger =
  S(R)^{-1}_{\alpha\beta}\, \bigl(\widetilde{D}^{(p)}_{R\hat{j}}
  \widetilde{\psi}(Rx)\bigr)^A_{a\beta},& \label{eq:transform}\\
  &U_R \bigl(\widetilde{\overline{\psi}}(x)\widetilde{D}^{(p)\dagger}_j
 \bigr)^A_{a\alpha}\! U_R^\dagger =
   \bigl(\widetilde{\overline{\psi}}(Rx)
   \widetilde{D}^{(p)\dagger}_{R\hat{j}}\bigr)^A_{a\beta}\, S(R)_{\beta\alpha}, &
\end{eqnarray}
where the transformation matrices for spatial inversion $I_s$
and proper rotations $C_{nj}$ through angle $2\pi/n$ about axis
$Oj$ are given by
\begin{eqnarray}
 S(C_{nj}) &=& \exp\Bigl(\textstyle\frac{1}{8}\omega_{\mu\nu}
 [\gamma_\mu,\gamma_\nu]  \Bigr),\\
S(I_s) &=& \gamma_4,
\end{eqnarray}
with $\omega_{kl}=-2\pi\varepsilon_{jkl}/n$ and $\omega_{4k}=\omega_{k4}=0$
($\omega_{\mu\nu}$ is an antisymmetric tensor which parametrizes rotations
and boosts). A rotation by $\pi/2$ about the $y$-axis is conventionally 
denoted by $C_{4y}$, and $C_{4z}$ denotes a rotation by $\pi/2$ about the 
$z$-axis.  These particular group elements are given by
\begin{equation}
 S(C_{4y})=\frac{1}{\sqrt{2}}(1+\gamma_1\gamma_3),\quad 
 S(C_{4z})=\frac{1}{\sqrt{2}}(1+\gamma_2\gamma_1).
\end{equation}
The allowed rotations on a three-dimensional spatially-isotropic cubic lattice
form the octahedral group $O$ which has 24 elements.  Inclusion of spatial
inversion yields the point group $O_h$ which has 48 elements occurring
in ten conjugacy classes. All elements of $O_h$ can be generated from 
appropriate products of only $C_{4y}$, $C_{4z}$, and $I_s$.

Charge conjugation is another symmetry of our theory.  Under charge
conjugation ${\cal C}$, the link variables $U\rightarrow U^\ast$ and
our basic building blocks transform according to
\begin{eqnarray}
  {\cal C}\ \bigl(\widetilde{D}^{(p)}_j
  \widetilde{\psi}(x)\bigr)^A_{a\alpha}\ {\cal C}^\dagger &=&
  \bigl(\widetilde{\overline{\psi}}(x)\widetilde{D}^{(p)\dagger}_j\bigr)^A_{a\beta}
 \  C_{\beta\alpha}^\dagger,\\
  {\cal C}\ \bigl(\widetilde{\overline{\psi}}(x)\widetilde{D}^{(p)\dagger}_j
  \bigr)^A_{a\alpha}\ {\cal C}^\dagger &=& -C_{\alpha\beta}
 \bigl(\widetilde{D}^{(p)}_j
  \widetilde{\psi}(x)\bigr)^A_{a\beta},
\end{eqnarray}
where the charge conjugation matrix $C$ must be unitary, antisymmetric, and satisfy
$C\gamma_\mu C^\dagger = -\gamma_\mu^T$.  Our choice for $C$ in the 
Dirac-Pauli representation is $C=\gamma_4\gamma_2$.

Operators which transform according to the irreducible representations of $O_h$
can then be constructed using the well-known group-theoretical projections
given in Eq.~(\ref{eq:project}).  Orthogonality
relations and hence, projection techniques, in group theory apply only to
single-valued irreducible representations.  However, the fermionic
representations are double-valued representations of $O_h$.  The commonly-used
trick to circumvent this difficulty is to exploit the equivalence of the
double-valued irreps of $O_h$ with the extra single-valued irreps of
the so-called {\em double point group} $O_h^D$.  This group is formed by
introducing a new element $\overline{E}$ which represents a rotation by
an angle $2\pi$ about any axis, such that $\overline{E}^2=E$ (the identity).
By including such an element, the total number of elements in $O_h^D$ is
double the number of elements in $O_h$.  The 96 elements of $O_h^D$ occur
in sixteen conjugacy classes.  

Since baryons are fermions, we need only be concerned with the six double-valued
irreps of $O_h$.  There are four two-dimensional irreps
$G_{1g}, G_{1u}, G_{2g}$, and $G_{2u}$, and two four-dimensional irreps
$H_g$ and $H_u$.  The subscript $g$ refers to even parity states, whereas the
subscript $u$ refers to odd parity states. The irreps $G_{1g}$ and $G_{1u}$ 
contain the spin-1/2 states, spin-3/2 states reside in the $H_g$ and $H_u$,
and two of the spin projections of the spin-5/2 states occur in the $G_{2g}$
and $G_{2u}$ irreps, while the remaining four projections reside in the
$H_g$ and $H_u$ irreps.  The spin content of each $O_h$ irrep in the continuum limit
is summarized in Table~\ref{tab:spin}.  This table lists the number of times
that each of the $O_h$ irreps occurs in the 
$J=\frac{1}{2},\frac{3}{2},\frac{5}{2},\cdots$ representations
of $SU(2)$ subduced to $O_h$. 

To carry out the projections in Eq.~(\ref{eq:project}),
explicit representation matrices are needed. Our choice of representation matrices
is summarized in Table~\ref{tab:ODreps}.  Matrices for only the
group elements $C_{4y},C_{4z}$, and $I_s$ are given in Table~\ref{tab:ODreps}
since the representation matrices for all other group elements can be obtained
by suitable multiplications of the matrices for the three generating elements.
For baryons, the representation matrix for $\overline{E}$ in each of the $O_h^D$
extra irreps is $-1$ times the identity matrix.

\begin{table}
\caption[tabODrep]{Our choice of the representation matrices for the double-valued 
 irreps of $O_h$.  The $G_{1u},G_{2u},H_u$ matrices for the rotations $C_{4y},C_{4z}$
  are the same as the $G_{1g},G_{2g},H_g$ matrices, respectively, given below.
  Each of the $G_{1g},G_{2g},H_g$ matrices for spatial inversion $I_s$ is the identity
  matrix, whereas each of the $G_{1u},G_{2u},H_u$ matrices for $I_s$ is $-1$ times the identity
  matrix. The matrices for all other group elements can be obtained from appropriate
  multiplications of the $C_{4y},C_{4z}$, and $I_s$ matrices.
 \label{tab:ODreps}}
\begin{ruledtabular}
\begin{tabular}{ccc}
 $\Lambda$ & $\Gamma^{(\Lambda)}(C_{4y})$
 & $\Gamma^{(\Lambda)}(C_{4z})$ 
 \\ \hline
$G_{1g}$  & \rule[-3ex]{0mm}{8ex}
$\displaystyle\frac{1}{\sqrt{2}}\!\left[\begin{array}{rr}
 1 & -1 \\ 1 &  1
\end{array}\right]$ &
$\displaystyle\frac{1}{\sqrt{2}}\!\left[\begin{array}{cc}
  1\!-\!i & 0 \\ 0 & 1\!+\!i 
\end{array}\right] $ \\
$G_{2g}$   &\rule[-3ex]{0mm}{8ex}
$\displaystyle\frac{-1}{\sqrt{2}}\!\left[\begin{array}{rr}
 1 & -1 \\ 1 &  1
\end{array}\right]$ &
 $\displaystyle\frac{-1}{\sqrt{2}}\!\left[\begin{array}{cc}
  1\!-\!i & 0 \\ 0 & 1\!+\!i 
\end{array}\right]$ \\
$H_g$  &\rule[-7ex]{0mm}{15ex}
$\displaystyle\!\!\frac{1}{2\sqrt{2}}\!\!\left[\begin{array}{rrrr}
  \!1 & \!\!-\sqrt{3} & \sqrt{3} & -1 \\
  \!\!\sqrt{3} & -1 & -1 &  \sqrt{3} \\
  \!\!\sqrt{3} &  1 & -1 & \!\!-\sqrt{3} \\
  1 &  \sqrt{3} & \sqrt{3} &  1 \end{array}\right] $
& $\displaystyle\!\!\frac{1}{\sqrt{2}}\!\!\left[\begin{array}{cccc}
   \!\!-1\!-\!i\!\! & 0 & 0 & 0 \\
   0 & \!\!1\!-\!i\!\! & 0 & 0 \\
   0 & 0 & \!\!1\!+\!i\!\! & 0 \\
   0 & 0 & 0 & \!\!-1\!+\!i\!\! \end{array}\right]$ 
\end{tabular}
\end{ruledtabular}
\end{table}

Note that Table~\ref{tab:spin} is the key to identifying the continuum-limit
spin $J$ corresponding to the masses extracted in our Monte Carlo calculations.  
For example, to identify an even parity baryon as having $J=1/2$, a level must
be observed in the $G_{1g}$ channel, and there must be no degenerate partners
in either of the $G_{2g}$ or $H_g$.  A level observed in the $H_g$ channel
with no degenerate partners in the $G_{1g}$ and $G_{2g}$ channels (in the
continuum limit) is a $J=3/2$ state.  Degenerate partners observed in the
$G_{2g}$ and $H_g$ channels with no partner in the $G_{1g}$ channel indicates
a $J=5/2$ baryon.  In other words, Table~\ref{tab:spin} details the patterns
of continuum-limit degeneracies corresponding to each half-integral $J$ value.

Our operators are constructed using fermion fields $\psi(x)$ which annihilate
a quark and create an antiquark.  Hence, each of our baryon operators 
annihilates a three-quark system of a given parity $P$ and creates
a three-antiquark system of the {\em same} parity $P$.  This means
that in the baryon propagator, a baryon of parity $P$ propagates forward in
time while an antibaryon of parity $P$ propagates backwards in time.  Unlike
boson fields, a fermion and its antifermion have opposite intrinsic parities,
so that the antibaryon propagating backwards in time is the antiparticle of
the \textit{parity partner} of the baryon propagating forwards in time.  Since
chiral symmetry is spontaneously broken, the masses of parity partners may differ.
The forward propagating baryon will have a mass different from that of
the antibaryon propagating backwards in time.  If the even and odd parity baryon
operators are carefully designed with respect to one another, it is possible to
arrange a definite relationship between the correlation matrix elements of
one parity for $t>0$ and the opposite-parity matrix elements
for $t<0$, allowing an increase in statistics.  Our operators are designed to
take advantage of this symmetry (see below). 

Our construction of the irreducible $B^{\Lambda\lambda F}_i(t)$ baryon operators
is done in the following sequence of steps.

(a) A set of $M_B$ linearly-independent elemental operators
$B_j^F(t)$ that transform among one another under $O_h^D$ is identified
with the help of the computer software mentioned earlier.  This is done
by starting with all possible operators of a given type (single-site,
singly-displaced, and so on), then using the \textsc{Maple} program to
detect dependencies between the operators.  To find such dependencies,
the computer program expresses each $B_j^F(t)$ operator as a sum of products
of Grassmann fields (or gauge-covariantly displaced Grassmann fields)
with explicit color, flavor, and spin indices: $B_j^F=\sum_k g^F_{jk} O_k$.
The coefficients $g^F_{jk}$ form a $M_B\times M_O$ matrix, where $M_O$
is the total number of $O_k$ operators encountered, which is much larger
than $M_B$.  Since each $O_k$ operator is a single product of three 
displaced Grassmann fields, these operators are linearly independent,
so the linear independence of the $B^F_j$ operators boils down to the linear
independence of the rows of the $g^F_{jk}$ matrix.  In this way, the set
of all possible operators of a given type is easily reduced to a set
containing only linearly-independent operators.

(b) We obtain the $M_B\times M_B$ representation matrices $W_{ij}(R)$ which
satisfy
\begin{equation}
  U_R\ B_i^F(t)\ U_R^\dagger = \sum_{j=1}^{M_B} B_j^F(t)\ W_{ji}(R).
\label{eq:BWrep}
\end{equation}
Our \textsc{Maple} program determines the $i$-th column of the $W_{ji}(R)$
matrix for a given symmetry transformation $R$ as follows.  First, the 
$B_i^F(t)$ operator is expressed as a sum of products of displaced
Grassmann fields with explicit color, flavor, and spin indices as in
the previous step:  $B_i^F=\sum_k g^F_{ik} O_k$.  Next, the $R$ symmetry
transformation is applied to the displaced Grassmann fields in each $O_k$
term in this sum of products using Eq.~(\ref{eq:transform}).  The resulting
sum of terms $U_R\ B_i^F(t)\ U_R^\dagger=\sum_k h^F_{ik} O_k$ is then
expressed as a linear superposition of the original $M_B$ operators 
using the Moore-Penrose pseudoinverse\cite{penrose} of the $g^F_{ik}$ matrix.
If the transformed operator contains Grassmann products which are not in the
original set of $O_k$ operators, then this signals that the starting 
basis of $B_j^F$ operators is incomplete, but with our method in the previous
step of choosing linearly-independent operators, this did not occur.
In this way, the $W_{ji}(R)$ matrices for the generating group elements
$C_{4y}$, $C_{4z}$, and $I_s$ are obtained.  The matrices for all other
group elements are then determined by appropriate products of these
three matrices.  At this point, we have a set of $M_B$ operators $B_i^F(t)$
which form the basis of a reducible representation given by the $W_{ij}(R)$
matrices.  Our remaining task is to find a change of basis such that the
resulting representation matrices are block-diagonal with the blocks given
by the irreducible representations of $O_h$.  

(c) Since the resulting $W_{ji}(R)$ matrices may not be unitary, we compute
the hermitian metric matrix $M$
\begin{equation}
 M_{ij} = \frac{1}{g_{O_h^D}}\sum_{R\in O_h^D} \sum_{k=1}^{M_B}
   W_{ki}(R)^\ast W_{kj}(R).
\end{equation}
This matrix will be needed in a later step where it will facilitate
the full block-diagonalization of the $W_{ij}(R)$ matrices.

(d) For each \textit{even}-parity irrep $\Lambda$, we compute the large
$M_B\!\times\! M_B$ projection matrix for row $\lambda=1$:
\begin{equation}
  P^{\Lambda\lambda F}_{ij}\! =\!
 \frac{d_\Lambda}{g_{O_h^D}}\sum_{R\in O_h^D} 
  \left[\Gamma^{(\Lambda)}_{\lambda\lambda}(R) W_{ji}(R)
 \right]_{\lambda\!=\!1}.
\end{equation}
This is one of the most important steps in our operator construction.  Applying
the group-theoretical projection of Eq.~(\ref{eq:project}) to the operator
$B_i^F$, then utilizing Eq.~(\ref{eq:BWrep}), produces a new operator
${\cal B}_i^{\Lambda\lambda F}(t)=\sum_j P^{\Lambda\lambda F}_{ij}B_j^F(t)$
which resides in the subspace of operators which transform according
to the row $\lambda=1$ of the given irrep $\Lambda$.  In other words,
the $i$-th row of the projection matrix $P$ contains the superposition
coefficients of the projected ${\cal B}^{\Lambda\lambda F}_i$ operator.  

(e) Although group theory guarantees that the resulting projected
${\cal B}^{\Lambda\lambda F}_i$ operators reside in the subspace associated
with row $\lambda$ of the $\Lambda$ irrep,
it does not guarantee that all of the resulting operators are linearly
independent.  The maximum number of independent operators in the
projected subspace is given by the rank $r$ of the projection matrix. 
Hence, the next step is to form $r$ superpositions of the
${\cal B}^{\Lambda\lambda F}_i$ operators such that the resulting $r$
operators are linearly independent.  The choice of these operators
is not unique.  In practice, these linear combinations are obtained using
the well-known Gram-Schmidt procedure, but with a modified inner product
to incorporate the metric matrix $M$.  The use of the metric matrix $M$
ensures full block-diagonalization of the original $W_{ij}(R)$ matrices.  Using
such a procedure, the final operators, expressed in terms of the original
set of $B_i^F$ operators by
\begin{equation}
 B^{\Lambda\lambda F}_i(t) = \sum_{j=1}^{M_B}\ c^{\Lambda\lambda F}_{ij}
 \ B_j^F(t), \quad  (\lambda=1)
\end{equation}
have superposition coefficients $c^{\Lambda\lambda F}_{ij}$ that satisfy
\begin{equation}
 \sum_{k,l=1}^{M_B} c^{\Lambda\lambda F\ast}_{ik}
 \ M_{kl}\ c^{\Lambda\lambda F}_{jl} = \delta_{ij},
 \ (i=1 \dots  r).
\end{equation}

(f) For each of the $r$ operators $B^{\Lambda\lambda F}_i(t)$ in
the first row  $\lambda=1$, we obtain partner operators in all other 
rows $\mu>1$ using
\begin{equation}
 c^{\Lambda\mu F}_{ik}
  = \sum_{j=1}^{M_B} c^{\Lambda\lambda F}_{ij}
  \frac{d_\Lambda}{g_{O_h^D}}\sum_{R\in O_h^D} 
  \Gamma^{(\Lambda)}_{\mu\lambda}(R)
  \ W_{kj}(R).
\end{equation}
The use of operators belonging to other rows will be important for
increasing the statistics of our Monte Carlo calculations, as will be
discussed below.

(g) Although the \textit{odd}-parity operators can be obtained using the
same procedure described above, we instead utilize charge conjugation 
to construct the odd-parity operators. Consider the correlation matrix element
of even-parity operators for $t\geq 0$.  Suppressing flavor and displacement indices,
one sees that invariance under charge conjugation implies that
\begin{eqnarray*}
 C_{ij}(t) 
&=& c^{(i)}_{\alpha\beta\gamma}c^{(j)\ast}_{\overline{\alpha}
\overline{\beta}\overline{\gamma}}
\langle 0\vert\ \Phi_{\alpha\beta\gamma}(t)
\overline{\Phi}_{\overline{\alpha}
\overline{\beta}\overline{\gamma}}(0)
\ \vert 0\rangle,\\
&=& c^{(i)}_{\alpha\beta\gamma}c^{(j)\ast}_{\overline{\alpha}
\overline{\beta}\overline{\gamma}}
\langle 0\vert\ {\cal C}^\dagger {\cal C}\Phi_{\alpha\beta\gamma}(t)
{\cal C}^\dagger {\cal C}\overline{\Phi}_{\overline{\alpha}
\overline{\beta}\overline{\gamma}}(0)
{\cal C}^\dagger {\cal C}\ \vert 0\rangle,\\
&=& c^{(i)}_{\alpha\beta\gamma}c^{(j)\ast}_{\overline{\alpha}
\overline{\beta}\overline{\gamma}}
\langle 0\vert\ \overline{\Phi}_{\alpha^\prime\beta^\prime\gamma^\prime}(t)
\Phi_{\overline{\alpha}^\prime
\overline{\beta}^\prime\overline{\gamma}^\prime}(0)
\ \vert 0\rangle\\
&&\times \gamma^2_{\alpha^\prime\alpha}\gamma^2_{\beta^\prime\beta}
\gamma^2_{\gamma^\prime\gamma}\gamma^2_{\overline{\alpha}^\prime\overline{\alpha}}
\gamma^2_{\overline{\beta}^\prime\overline{\beta}}
\gamma^2_{\overline{\gamma}^\prime\overline{\gamma}},\\
&=& c^{(i)}_{\alpha\beta\gamma}c^{(j)\ast}_{\overline{\alpha}
\overline{\beta}\overline{\gamma}}
\langle 0\vert\ \overline{\Phi}_{\alpha^\prime\beta^\prime\gamma^\prime}(0)
\Phi_{\overline{\alpha}^\prime
\overline{\beta}^\prime\overline{\gamma}^\prime}(-t)
\ \vert 0\rangle\\ &&\times \gamma^2_{\alpha^\prime\alpha}\gamma^2_{\beta^\prime\beta}
\gamma^2_{\gamma^\prime\gamma}\gamma^2_{\overline{\alpha}^\prime\overline{\alpha}}
\gamma^2_{\overline{\beta}^\prime\overline{\beta}}
\gamma^2_{\overline{\gamma}^\prime\overline{\gamma}},
\end{eqnarray*}
using invariance under time translations of the above expectation value
and invariance of the vacuum under charge conjugation.  The last line above represents
the correlation of odd-parity operators propagating temporally backwards.  Hence,
for a given even-parity operator $B^g_i(t)$, we can define an odd-parity
operator $B^u_i(t)$ by rotating the three Dirac indices using the $\gamma_2$ matrix and
replacing the expansion coefficients by their complex conjugates
such that the correlation matrices of the even and odd parity operators
are related by
\begin{eqnarray}
&C^{G_{1g}}_{ij}(t)=-C^{G_{1u}}_{ij}(-t)^\ast,\quad 
C^{G_{2g}}_{ij}(t)=-C^{G_{2u}}_{ij}(-t)^\ast,&\nonumber\\ 
& C^{H_{g}}_{ij}(t)=-C^{H_{u}}_{ij}(-t)^\ast.&
\label{eq:parityflip}
\end{eqnarray}
For
\begin{table}[t]
\caption[tabC]{The single-site $\Delta^{++}$
 operators which transform irreducibly under the symmetry group
 of the spatial lattice, defining 
 $\Delta_{\alpha\beta\gamma}=\Phi^{uuu}_{\alpha\beta\gamma;000}$
 (see Eq.~(\protect\ref{eq:giform})).
\label{tab:SSDelta}}
\begin{ruledtabular}
\begin{tabular}{c@{\hspace{2em}}cc@{\hspace{2em}}cc}
 Irrep & Row & Operator & Row & Operator\\ \hline 
 $G_{1g}$ & 1 &  $\Delta_{134}-\Delta_{233} $ 
          & 2 &  $\Delta_{144}-\Delta_{234} $ \\ 
 $G_{1u}$ & 1 &  $\Delta_{123}-\Delta_{114} $ 
          & 2 &  $\Delta_{223}-\Delta_{124} $ \\ 
 $H_{g}$ & 1 &  $\Delta_{111} $ 
         & 2 &  $\sqrt{3}\,\Delta_{112} $ \\
 $H_{g}$ & 3 &  $\sqrt{3}\,\Delta_{122} $ 
         & 4 &  $\Delta_{222} $ \\ 
 $H_{g}$ & 1 &  $\sqrt{3}\,\Delta_{133} $ 
         & 2 &  $2\Delta_{134}+\Delta_{233} $ \\
 $H_{g}$ & 3 &  $\Delta_{144}+2\Delta_{234} $ 
         & 4 &  $\sqrt{3}\,\Delta_{244} $ \\ 
 $H_{u}$ & 1 &  $\Delta_{333} $ 
         & 2 &  $\sqrt{3}\,\Delta_{334} $ \\
 $H_{u}$ & 3 &  $\sqrt{3}\,\Delta_{344} $ 
         & 4 &  $\Delta_{444} $ \\ 
 $H_{u}$ & 1 &  $\sqrt{3}\,\Delta_{113} $ 
         & 2 &  $\Delta_{114}+2\Delta_{123} $ \\
 $H_{u}$ & 3 &  $2\Delta_{124}+\Delta_{223} $ 
         & 4 &  $\sqrt{3}\,\Delta_{224} $ 
\end{tabular}
\end{ruledtabular}
\end{table}
\begin{table}[b]
\caption[tabC]{The single-site $\Lambda^0$
 operators which transform irreducibly under the symmetry group
 of the spatial lattice, defining 
 $\Lambda_{\alpha\beta\gamma}=\Phi^{uds}_{\alpha\beta\gamma;000}
  -\Phi^{dus}_{\alpha\beta\gamma;000}$
 (see Eq.~(\protect\ref{eq:giform})).
\label{tab:SSLambda}}
\begin{ruledtabular}
\begin{tabular}{c@{\hspace{1em}}cc@{\hspace{1em}}cc}
 Irrep & Row & Operator & Row & Operator\\ \hline 
 $G_{1g}$ & 1 &  $\Lambda_{121} $  
          & 2 &  $\Lambda_{122} $ \\
 $G_{1g}$ & 1 &  $\Lambda_{341} $ 
          & 2 &  $\Lambda_{342} $ \\
 $G_{1g}$ & 1 &  $\Lambda_{134}-\Lambda_{143} $ 
          & 2 &  $\Lambda_{234}-\Lambda_{243} $ \\
 $G_{1g}$ & 1 &  $\Lambda_{134}\!+\!\Lambda_{143}\!-\!2\Lambda_{233} $
          & 2 &  $2\Lambda_{144}\!-\!\Lambda_{234}\!-\!\Lambda_{243} $ \\
 $G_{1u}$ & 1 &  $\Lambda_{343} $ 
          & 2 &  $\Lambda_{344} $ \\
 $G_{1u}$ & 1 &  $\Lambda_{123} $ 
          & 2 &  $\Lambda_{124} $ \\
 $G_{1u}$ & 1 &  $\Lambda_{231}-\Lambda_{132} $ 
          & 2 &  $\Lambda_{241}-\Lambda_{142} $ \\
 $G_{1u}$ & 1 &  $2\Lambda_{141}\!-\!\Lambda_{132}\!-\!\Lambda_{231} $ 
          & 2 &  $\Lambda_{142}\!-\!2\Lambda_{232}\!+\!\Lambda_{241} $ \\
 $H_{g}$ & 1 &  $\sqrt{3}\,\Lambda_{133} $ 
         & 2 &  $\Lambda_{134}\!+\!\Lambda_{143}\!+\!\Lambda_{233} $ \\
 $H_{g}$ & 3 &  $\Lambda_{144}\!+\!\Lambda_{234}\!+\!\Lambda_{243} $ 
         & 4 &  $\sqrt{3}\,\Lambda_{244} $ \\
 $H_{u}$ & 1 &  $-\sqrt{3}\,\Lambda_{131} $ 
         & 2 &  $\!-\!\Lambda_{132}\!-\!\Lambda_{141}\!-\!\Lambda_{231} $ \\
 $H_{u}$ & 3 &  $\!-\!\Lambda_{142}\!-\!\Lambda_{232}\!-\!\Lambda_{241} $ 
         & 4 &  $-\sqrt{3}\,\Lambda_{242} $
\end{tabular}
\end{ruledtabular}
\end{table}
 a lattice of $N_t$ sites in the time direction with periodic $(\eta_t=1)$
or antiperiodic $(\eta_t=-1)$ boundary conditions, this means that
\begin{equation}
C^{G_{1g}}_{ij}(t)=-\eta_t\ C^{G_{1u}}_{ij}(N_t-t)^\ast,
\end{equation}
and similarly for the other irreps.
This allows us to appropriately average over forward and backward temporal
propagations for increased statistics.

\begin{table}[t]
\caption[tabC]{The single-site $\Sigma^+$ 
 operators which transform irreducibly under the symmetry group
 of the spatial lattice, defining 
 $\Sigma_{\alpha\beta\gamma}=\Phi^{uus}_{\alpha\beta\gamma;000}$
 (see Eq.~(\protect\ref{eq:giform})).
\label{tab:SSSigma}}
\begin{ruledtabular}
\begin{tabular}[t]{c@{\hspace{1em}}cc@{\hspace{1em}}cc}
 Irrep & Row & Operator & Row & Operator\\ \hline 
 $G_{1g}$ & 1 &  $\Sigma_{112}-\Sigma_{121} $ 
          & 2 &  $\Sigma_{122}-\Sigma_{221} $ \\ 
 $G_{1g}$ & 1 &  $\Sigma_{134}-\Sigma_{143} $ 
          & 2 &  $\Sigma_{234}-\Sigma_{243} $ \\ 
 $G_{1g}$ & 1 &  $\Sigma_{332}-\Sigma_{341} $ 
          & 2 &  $\Sigma_{342}-\Sigma_{441} $ \\ 
 $G_{1g}$ & 1 &  $\Sigma_{134}\!+\!\Sigma_{143}\!-\!2\Sigma_{233} $ 
          & 2 &  $2\Sigma_{144}\!-\!\Sigma_{234}\!-\!\Sigma_{243} $ \\ 
 $G_{1u}$ & 1 &  $\Sigma_{334}-\Sigma_{343} $ 
          & 2 &  $\Sigma_{344}-\Sigma_{443} $ \\ 
 $G_{1u}$ & 1 &  $\Sigma_{132}-\Sigma_{231} $ 
          & 2 &  $\Sigma_{142}-\Sigma_{241} $ \\ 
 $G_{1u}$ & 1 &  $\Sigma_{114}-\Sigma_{123} $ 
          & 2 &  $\Sigma_{124}-\Sigma_{223} $ \\ 
 $G_{1u}$ & 1 &  $\Sigma_{132}\!-\!2\Sigma_{141}\!+\!\Sigma_{231} $ 
          & 2 &  $2\Sigma_{232}\!-\!\Sigma_{241}\!-\!\Sigma_{142} $ \\ 
 $H_{g}$ & 1 &  $\sqrt{3}\,\Sigma_{111} $ 
         & 2 &  $\Sigma_{112}+2\Sigma_{121} $ \\
 $H_{g}$ & 3 &  $2\Sigma_{122}+\Sigma_{221} $ 
         & 4 &  $\sqrt{3}\,\Sigma_{222} $ \\ 
 $H_{g}$ & 1 &  $\sqrt{3}\,\Sigma_{133} $ 
         & 2 &  $\Sigma_{134}\!+\!\Sigma_{143}\!+\!\Sigma_{233} $ \\
 $H_{g}$ & 3 &  $\Sigma_{144}\!+\!\Sigma_{234}\!+\!\Sigma_{243} $ 
         & 4 &  $\sqrt{3}\,\Sigma_{244} $ \\ 
 $H_{g}$ & 1 &  $\sqrt{3}\,\Sigma_{331} $ 
         & 2 &  $\Sigma_{332}+2\Sigma_{341} $ \\
 $H_{g}$ & 3 &  $2\Sigma_{342}+\Sigma_{441} $ 
         & 4 &  $\sqrt{3}\,\Sigma_{442} $ \\ 
 $H_{u}$ & 1 &  $\sqrt{3}\,\Sigma_{333} $ 
         & 2 &  $\Sigma_{334}+2\Sigma_{343} $ \\
 $H_{u}$ & 3 &  $2\Sigma_{344}+\Sigma_{443} $ 
         & 4 &  $\sqrt{3}\,\Sigma_{444} $ \\ 
 $H_{u}$ & 1 &  $\sqrt{3}\,\Sigma_{131} $ 
         & 2 &  $\Sigma_{132}\!+\!\Sigma_{141}\!+\!\Sigma_{231} $ \\
 $H_{u}$ & 3 &  $\Sigma_{142}\!+\!\Sigma_{232}\!+\!\Sigma_{241} $ 
         & 4 &  $\sqrt{3}\,\Sigma_{242} $ \\ 
 $H_{u}$ & 1 &  $\sqrt{3}\,\Sigma_{113} $ 
         & 2 &  $\Sigma_{114}+2\Sigma_{123} $ \\
 $H_{u}$ & 3 &  $2\Sigma_{124}+\Sigma_{223} $ 
         & 4 &  $\sqrt{3}\,\Sigma_{224} $ 
\end{tabular}
\end{ruledtabular}
\end{table}

\begin{table}[b]
\caption[tabC]{The single-site $N^+$ 
 operators which transform irreducibly under the symmetry group
 of the spatial lattice, defining 
 $N_{\alpha\beta\gamma}=\Phi^{uud}_{\alpha\beta\gamma;000}
-\Phi^{duu}_{\alpha\beta\gamma;000}$ (see Eq.~(\protect\ref{eq:giform})).
\label{tab:SSNucleon}}
\begin{ruledtabular}
\begin{tabular}{c@{\hspace{1em}}cc@{\hspace{1em}}cc}
 Irrep & Row & Operator & Row & Operator\\ \hline 
 $G_{1g}$ & 1 &  $N_{211} $ 
          & 2 &  $N_{221} $ \\
 $G_{1g}$ & 1 &  $N_{413} $ 
          & 2 &  $N_{423} $ \\
 $G_{1g}$ & 1 &  $2N_{332}\!+\!N_{413}\!-\!2N_{431}\!\!\! $
          & 2 &  $2N_{432}\!-\!2N_{441}\!-\!N_{423} $ \\ 
 $G_{1u}$ & 1 &  $N_{433} $ 
          & 2 &  $N_{443} $ \\ 
 $G_{1u}$ & 1 &  $N_{321}-N_{312} $ 
          & 2 &  $N_{421}-N_{412} $ \\ 
 $G_{1u}$ & 1 &  $N_{312}\!+\!N_{321}\!-\!2N_{411} $ 
          & 2 &  $2N_{322}\!-\!N_{412}\!-\!N_{421} $ \\ 
 $H_{g}$ & 1 &  $\sqrt{3}\,N_{331} $
         & 2 &  $N_{332}\!-\!N_{413}\!+\!2N_{431} $ \\
 $H_{g}$ & 3 &  $2N_{432}\!+\!N_{441}\!-\!N_{423} $
         & 4 &  $\sqrt{3}\,N_{442} $ \\ 
 $H_{u}$ & 1 &  $-\sqrt{3}\,N_{311} $ 
         & 2 &  $\!-\!N_{312}\!-\!N_{321}\!-\!N_{411} $ \\
 $H_{u}$ & 3 &  $\!-\!N_{322}\!-\!N_{412}\!-\!N_{421} $ 
         & 4 &  $-\sqrt{3}\,N_{422} $ 
\end{tabular}
\end{ruledtabular}
\end{table}

\begin{table*}[t]
\caption[delproj]{The numbers of operators of each type which project
into each row of the $G_{1g}, G_{2g},$ and $H_g$ irreps for the 
$\Delta^{++}$, $\Sigma^+, N^+,$ and $\Lambda^0$ baryons.  The numbers
for the $G_{1u}, G_{2u},$ and $H_u$ irreps are the same as for the
$G_{1g}, G_{2g},$ and $H_g$, respectively.
\label{tab:embeddings}}
\begin{ruledtabular}
\begin{tabular}{lrrr@{\hspace{3em}}rrr@{\hspace{3em}}rrr@{\hspace{3em}}rrr} 
 & \multicolumn{3}{c@{\hspace{3em}}}{$\Delta^{++}$} &  \multicolumn{3}{c@{\hspace{3em}}}{$\Sigma^+$} 
 & \multicolumn{3}{c@{\hspace{3em}}}{$N^+$} & \multicolumn{3}{c}{$\Lambda^0$} \\
 Operator type      &  $G_{1g}$  &  $G_{2g}$  &   $H_g$  &  $G_{1g}$  &  $G_{2g}$  &   $H_g$ &  $G_{1g}$  &  $G_{2g}$  &   $H_g$ &  $G_{1g}$  &  $G_{2g}$  &   $H_g$\\ \hline
 single-site        &     1      &     0      &     2    &     4      &     0      &     3   &     3      &     0      &     1   &     4      &     0      &     1  \\
 singly-displaced   &    14      &     6      &    20    &    38      &    14      &    52   &    24      &     8      &    32   &    34      &    10      &    44  \\
 doubly-displaced-I &    12      &     4      &    16    &    36      &    12      &    48   &    24      &     8      &    32   &    36      &    12      &    48  \\
 doubly-displaced-L &    32      &    32      &    64    &    96      &    96      &   192   &    64      &    64      &   128   &    96      &    96      &   192  \\
 triply-displaced-T &    32      &    32      &    64    &    96      &    96      &   192   &    64      &    64      &   128   &    96      &    96      &   192  \\
 triply-displaced-O &    20      &    20      &    44    &    64      &    64      &   128   &    44      &    44      &    84   &    64      &    64      &   128  
\end{tabular}
\end{ruledtabular}
\end{table*}

In the absence of any external applied fields, the energies of the baryons
do not depend on the row $\lambda$ of a given irrep $\Lambda$, so we can 
increase statistics by averaging over all rows. The construction of the operators
in the different rows of the irreps as described above leads to correlation
matrices which satisfy
$ C^{\Lambda\lambda F}_{ij}(t) =C^{\Lambda\mu F}_{ij}(t),$
for all rows $\lambda,\mu$.  Hence, the correlation matrix elements themselves
can be averaged over rows.

The single-site operators produced by the above procedure are presented in
Tables~\ref{tab:SSDelta}-\ref{tab:SSNucleon}.  It is not possible to list
all of the singly-displaced, doubly-displaced, and triply-displaced operators
in this paper.  Instead, we simply list the numbers of operators of each type
which project into each row of the irreps for the $\Delta^{++}$, $\Sigma^+, 
N^+,$ and $\Lambda^0$ baryons in Table~\ref{tab:embeddings}.  Further details
about these operators are available upon request.

Note that we have not yet attempted to remove redundant operators related to 
others by a total lattice derivative, and that there may exist relationships 
between the correlation matrix elements of these operators from the equations
of motion.  Such relationships can be easily identified using a singular value
decomposition of the correlation matrices in small-lattice low-statistics 
Monte Carlo calculations.  Since the Schwinger-Dyson equations may relate
operators of different types, such as singly-displaced operators
with single-site operators, identifying these relationships must be done
at a later stage in the calculations.

\section{Baryon propagators}
\label{sec:baryonprop}

To extract the baryon masses, we need to compute the correlations
$C_{ij}^{\Lambda\lambda F}(t)=\langle 0\vert\ T B_i^{\Lambda\lambda F}(t)
 \overline{B}_j^{\Lambda\lambda F}(0)\ \vert 0\rangle,$
using the operators constructed as described above.  In this section,
several issues related to computing these correlation matrix elements are
discussed.  In particular, we discuss the use of symmetry to minimize
the number of quark-propagator sources, the use of gauge-invariant three-quark
propagators as an intermediate step in the baryon propagator determinations,
and detail the application of Wick's theorem.

The baryon propagators may be expressed in terms of the correlations of
the elemental operators by
\begin{equation}
 C_{ij}^{\Lambda\lambda F}(t)=\sum_{k,\ l=1}^{M_B}
 c_{ik}^{\Lambda\lambda  F} c_{jl}^{\Lambda\lambda F\ast}
 \ \langle 0\vert\ T B_k^F(t)
 \overline{B}_l^F(0)\ \vert 0\rangle.
\end{equation}
Note that correlations between operators in different rows of the same irrep vanish.
Since the number of elemental operators is large and the quark propagators
are rather expensive to compute, it is very important to use symmetry
to reduce the number of quark-propagator sources.  Given the invariance of the
vacuum and the unitarity of the symmetry transformation operators, we
know that
\begin{equation}\begin{array}{l}
 \langle 0\vert\ T B_k^F(t)
 \overline{B}_l^F(0)\ \vert 0\rangle \nonumber\\
=  \langle 0\vert\ T \ U_R B_k^F(t) U_R^\dagger
 \ U_R\overline{B}_l^F(0) U_R^\dagger\ \vert 0\rangle,\nonumber\\
=  \displaystyle\sum_{k^\prime,l^\prime=1}^{M_B}
  \!\! W_{k^\prime k}(R) W_{l^\prime l}(R)^\ast
 \langle 0\vert T B_{k^\prime}^F(t) \overline{B}_{l^\prime}^F(0)
 \vert 0\rangle,
\end{array}\end{equation}
for any group element $R$ of $O_h$.  Hence, for each source 
$\overline{B}_l^F(0)$, we can choose a group element $R_l$ such that we
minimize the total number of source elemental operators which must be considered. 
For example, consider the singly-displaced operators.  We can choose
an $R_l$ such that the displaced quark in the source is always displaced in
the $+z$ direction.  Similarly, a group element $R_l$ can always be chosen to
rotate each of the other types of operators into a specific orientation.

The coefficients $c^{\Lambda\lambda F}_{ij}$ in the baryon operators
involve only the Dirac spin components and the quark displacement
directions and are independent of the color indices and spatial sites.  Thus,
in calculating the baryon correlators, it is convenient to first calculate
gauge-invariant three-quark propagators in which all summations over color
indices and spatial sites have been done.  A three-quark propagator is 
defined by
\begin{eqnarray}
&&\widetilde{G}^{(ABC)(p\overline{p})}_{(\alpha i\vert\overline{\alpha}\overline{i})
(\beta j\vert\overline{\beta}\overline{j})
(\gamma k\vert\overline{\gamma}\overline{k})}(t)\nonumber\\
&=&\sum_{\bm{x}}
\varepsilon_{abc}\,\varepsilon_{\overline{a}\overline{b}\overline{c}}
\ \widetilde{Q}^{(A)}_{a\alpha i p\vert\overline{a}
   \overline{\alpha}\overline{i}\overline{p}}(\bm{x},t\vert\bm{x}_0,0)\nonumber\\
&\times&\widetilde{Q}^{(B)}_{b\beta j p\vert\overline{b}
   \overline{\beta}\overline{j}\overline{p}}(\bm{x},t\vert\bm{x}_0,0)
\ \widetilde{Q}^{(C)}_{c\gamma k p\vert\overline{c}
   \overline{\gamma}\overline{k}\overline{p}}(\bm{x},t\vert\bm{x}_0,0),
\quad \label{eq:threequarkprop}
\end{eqnarray}
where $\widetilde{Q}^{(A)}_{a\alpha i p\vert\overline{a}
   \overline{\alpha}\overline{i}\overline{p}}(\bm{x},t\vert\bm{x}_0,0)$
denotes the propagator for a single smeared quark field of flavor $A$ from
source site $\bm{x}_0$ at time $t=0$ to sink site $\bm{x}$ at time $t$.
At the sink, $a$ denotes color, $\alpha$ is the Dirac spin index, $i$ is
the displacement direction, and $p$ is the displacement length, and
similarly at the source for $\overline{a},\overline{\alpha},\overline{i},$
and $\overline{p}$, respectively.  Notice that the three-quark propagator is
symmetric under interchange of all indices associated with the same flavor.
As usual, translation invariance is invoked at the source so that summation
over spatial sites is done only at the sink.  These three-quark propagators
are computed for all possible values of the six Dirac spin indices.

Each baryon correlator is simply a linear superposition of elements of the
three-quark propagators.  These superposition coefficients are calculated
as follows: first, the baryon operators at the source and sink are
expressed in terms of the elemental operators; next, Wick's theorem is
applied to express the correlator as a large sum of three-quark propagator
components;  finally, symmetry operations are applied to minimize the number
of source orientations, and the results are averaged over the rows of the 
representations.  C++ code was written to perform these computations,
and the resulting superposition coefficients are stored in computer files
which are subsequently used as input to the Monte Carlo runs.

Wick's theorem is an important part of expressing the baryon correlators
in terms of the three-quark propagators.  To simplify the notation in the
following, let the indices $\mu,\nu,\tau$ each represent a Dirac spin
index and a displacement direction, and suppress the displacement lengths.
Define
$   \overline{c}^{(i)}_{\mu\nu\tau} 
 = c^{(i)\ast}_{\mu^\prime\nu^\prime\tau^\prime}
 \gamma^4_{\mu\mu^\prime} \gamma^4_{\nu\nu^\prime} \gamma^4_{\tau\tau^\prime},$
then the elements of the baryon correlation matrix in the $\Delta^{++}$
channel are given in terms of three-quark propagator components (before
source-minimizing rotations) by
\begin{eqnarray}
 C^{(\Delta)}_{ij}(t) &=& c^{(i)}_{\mu\nu\tau}
\, \overline{c}^{(j)}_{\overline{\mu}\overline{\nu}\overline{\tau}}
\Bigl\{
\,\widetilde{G}^{(uuu)}_{(\tau  \vert \overline{\mu})
(\nu   \vert \overline{\nu})
(\mu   \vert \overline{\tau})}(t) \nonumber\\
&+&\,\widetilde{G}^{(uuu)}_{ (\tau  \vert \overline{\mu})
(\nu   \vert \overline{\tau})
(\mu   \vert \overline{\nu})}(t) 
+\,\widetilde{G}^{(uuu)}_{ (\tau  \vert \overline{\nu})
(\nu   \vert \overline{\mu})
(\mu   \vert \overline{\tau})}(t)\nonumber\\
&+&\,\widetilde{G}^{(uuu)}_{ (\tau  \vert \overline{\nu})
(\nu   \vert \overline{\tau})
(\mu   \vert \overline{\mu})}(t)
+\,\widetilde{G}^{(uuu)}_{ (\tau  \vert \overline{\tau})
(\nu   \vert \overline{\nu})
(\mu   \vert \overline{\mu})}(t) \nonumber\\
&+&\,\widetilde{G}^{(uuu)}_{ (\tau  \vert \overline{\tau})
(\nu   \vert \overline{\mu})
(\mu   \vert \overline{\nu})}(t)
\Bigr\}.
\end{eqnarray}
The $N^+$ correlators are expressed in terms of components of
three-quark propagators by
\begin{eqnarray}
 C^{(N)}_{ij}(t) &=& c^{(i)}_{\mu\nu\tau}
\, \overline{c}^{(j)}_{\overline{\mu}\overline{\nu}\overline{\tau}}
\Bigl\{
\widetilde{G}^{(uud)}_{ (\mu  \vert \overline{\mu})
(\nu  \vert \overline{\nu})
(\tau \vert \overline{\tau})} \nonumber\\
&+&\widetilde{G}^{(uud)}_{ (\mu  \vert \overline{\nu})
(\nu  \vert \overline{\mu})
(\tau \vert \overline{\tau})} 
-\widetilde{G}^{(uud)}_{ (\mu  \vert \overline{\tau})
(\nu  \vert \overline{\nu})
(\tau \vert \overline{\mu})} \nonumber\\
&-&\widetilde{G}^{(uud)}_{ (\mu  \vert \overline{\nu})
 (\nu  \vert \overline{\tau})
 (\tau \vert \overline{\mu})}
-\widetilde{G}^{(uud)}_{ (\nu  \vert \overline{\nu}) 
 (\tau \vert \overline{\mu})
 (\mu  \vert \overline{\tau})}\nonumber\\
&-&\widetilde{G}^{(uud)}_{ (\nu  \vert \overline{\mu})
 (\tau \vert \overline{\nu}) 
 (\mu  \vert \overline{\tau})}
+\widetilde{G}^{(uud)}_{ (\tau \vert \overline{\tau})
 (\nu  \vert \overline{\nu})
 (\mu  \vert \overline{\mu})}\nonumber\\
&+&\widetilde{G}^{(uud)}_{ (\tau \vert \overline{\nu})
 (\nu  \vert \overline{\tau})
 (\mu  \vert \overline{\mu})} 
\Bigr\},
\end{eqnarray}
and for the $\Sigma^+$ and $\Lambda^0$ channels, one finds
\begin{eqnarray}
 C^{(\Sigma)}_{ij}(t) &=& c^{(i)}_{\mu\nu\tau}
\, \overline{c}^{(j)}_{\overline{\mu}\overline{\nu}\overline{\tau}}
\Bigl\{
\widetilde{G}^{(uus)}_{ (\mu  \vert \overline{\mu})
(\nu  \vert \overline{\nu})
(\tau \vert \overline{\tau})}(t) \nonumber\\
&+&\widetilde{G}^{(uus)}_{ (\mu  \vert \overline{\nu})
(\nu  \vert \overline{\mu})
(\tau \vert \overline{\tau})}(t)\Bigr\},\\
 C^{(\Lambda)}_{ij}(t) &=& c^{(i)}_{\mu\nu\tau}
\, \overline{c}^{(j)}_{\overline{\mu}\overline{\nu}\overline{\tau}}
\Bigl\{
\widetilde{G}^{(uds)}_{ (\mu  \vert \overline{\mu})
( \nu  \vert \overline{\nu})
( \tau \vert \overline{\tau})} \nonumber\\
&-&\widetilde{G}^{(uds)}_{ (\mu  \vert \overline{\nu})
( \nu  \vert \overline{\mu})
( \tau \vert \overline{\tau})}
-\widetilde{G}^{(uds)}_{ (\nu  \vert \overline{\mu})
( \mu  \vert \overline{\nu})
( \tau \vert \overline{\tau})} \nonumber\\
&+&\widetilde{G}^{(uds)}_{ (\nu  \vert \overline{\nu})
( \mu  \vert \overline{\mu})
( \tau \vert \overline{\tau})}
\Bigr\}.
\end{eqnarray}

\section{Conclusion}
\label{sec:conclude}

We plan to undertake a comprehensive study of the spectrum of QCD using Monte
Carlo computations.  Our first goal in this study is to calculate the
masses of as many low-lying hadron resonances as possible.  Successfully
extracting these masses will depend crucially on using carefully designed
spatially-extended hadronic operators.  In this first paper, the construction
of three-quark $\Delta, N, \Sigma, \Lambda, \Omega,\Xi$ baryon operators
using group-theoretical projections was detailed. The operators were assembled 
out of gauge-covariantly displaced quark fields and transform according to the
irreducible representations of the symmetry group of a spatial simple cubic
lattice.  Single-site, singly-displaced, doubly-displaced, and triply-displaced
three-quark operators were considered.  The guiding principles in devising our
operators were maximizing overlaps with the states of interest while minimizing
the number of quark-propagator sources.  Identifying the continuum-limit spins
$J$ of the states was addressed, and various issues related to computing the
correlation matrix elements of the baryon operators were discussed.

Due to the complexity of these calculations and the importance of providing checks
on our final results, we have been pursuing two different approaches
to constructing the baryon operators.  An alternative method of building the
baryon operators based on Clebsch-Gordan techniques will be presented 
elsewhere\cite{maryland}.  The construction of meson and multi-hadron operators
will be described in subsequent papers.

The Monte Carlo software to evaluate the correlation matrix elements of these
baryons operators has been written and thoroughly tested using a large variety
of checks, including comparison with known results in the case of a uniform
constant background gauge field.  This software uses the Chroma Software 
System for Lattice QCD\cite{chroma} with QDP++ and QMP, developed under
the Scientific Discovery through Advanced Computing initiative of the
U.S.~Department of Energy. Preliminary results have already been presented
in Refs.~\cite{basak1,basak2}. Although a very large number of baryon operators
have been devised, it is not our intent to evaluate correlation matrices using all of
these operators. Such calculations would not be feasible.  The next important step
in our study is to remove dynamically-redundant and ineffective operators using
low-statistics Monte Carlo calculations, with the goal of finding some
reasonably small subsets of operators adequate for extracting the low-lying
masses of interest.  Such calculations are currently in progress. 

\begin{acknowledgments}
This work was supported by the U.S.\ National Science Foundation 
through grants PHY-0354982 and PHY-0300065, and by 
the U.S.~Dept.\ of Energy under
contracts DE-AC05-84ER40150 and DE-FG02-93ER-40762.
\end{acknowledgments}

\bibliography{cited_refs}
\end{document}